%% file: Entanglement_resonances.tex
\documentclass{iopart}

\pdfoutput=1

\usepackage{graphics}      
\usepackage[pdftex]{graphicx,color}

\expandafter\let\csname equation*\endcsname\relax 
\expandafter\let\csname endequation*\endcsname\relax 
\usepackage{amsmath}

\usepackage{amssymb}
\usepackage{braket}
\usepackage{bbm}
\usepackage{color}

\usepackage{rotating} 
\usepackage{tabularx} 
\newcommand{\bfl}{\begin{flushleft}} 
\newcommand{\efl}{\end{flushleft}} 

\begin{document}

\title{Entanglement resonances of driven multi-partite quantum systems}

\author{Simeon Sauer$^1$, Florian Mintert$^{1,2}$, Clemens Gneiting$^1$, Andreas Buchleitner$^1$}
\address{$^1$ Physikalisches Institut, Albert-Ludwigs-Universit\"at, Hermann-Herder-Str. 3, D-79104 Freiburg, Germany}
\address{$^2$ Freiburg Institute for Advanced Studies, Albert-Ludwigs-Universit\"at Freiburg, Albertstra\ss e 19, D-79104 Freiburg, Germany}

\date{\today}

\begin{abstract}
We show how to create maximally entangled dressed states of a weakly interacting multi-partite quantum system by suitably tuning an external, periodic driving field. Floquet theory allows us to relate, in a transparent manner, the occurrence of \textit{entanglement resonances} to avoided crossings in the spectrum of quasi-energies, tantamount of well-defined conditions for the controlled, resonant interaction of particles. We demonstrate the universality of the phenomenon for periodically driven, weakly interacting two-level systems, by considering different interaction mechanisms and driving profiles. In particular, we show that entanglement resonances are a generic feature of driven, multi-partite systems, widely independent of the details of the interaction mechanism. Our results are therefore particularly relevant for experiments on interacting two-level systems, in which the microscopic realization of the inter-particle coupling is unknown.\end{abstract}


\input{Introduction}

\input{Theory}

\input{2Qubits}

\input{3Qubits}

\input{NQubits}

\input{Conclusion}

\appendix
\input{Appendix}

\bibliographystyle{iopart-num}
\bibliography{bibliography}

\end{document}

%% file: Introduction.tex

\section{Introduction}

Strongly entangled multi-partite states of a collection of quantum mechanical objects are a central resource in quantum information science, think, e.g., of one-way quantum computing \cite{Raussendorf:2001}. Beyond that, entangled many-particle states naturally occur in quantum phase transitions \cite{Osterloh:2002}, and in the condensed phase \cite{Ghosh:2003}, and recently they are also debated in the context of non-equilibrium processes in biological tissue \cite{Chang:2012}.

A viable way to achieve long-lasting entanglement in autonomous quantum systems is to adjust the system parameters such that one of the eigenstates becomes entangled, and then to prepare the system in this eigenstate \cite{Facchi:2011}. The appeal of this scheme is that, in the idealized situation of an isolated quantum system, the system remains in this eigenstate in the course of time, and hence preserves its entanglement. Even if the idealization of an isolated system is not justified, the coupling to a noisy environment, if not too strong, typically affects system eigenstates least \cite{Paz:1999}, rendering their entanglement inherently robust against decoherence. Accordingly, entanglement of eigenstates of autonomous quantum systems has been studied extensively in the literature, e.g., for spin chains \cite{Amico:2008}.

In this paper, we extend this scheme to \textit{periodically driven} quantum systems, since the application of external, time-periodic driving fields is often the most elementary way to control a quantum system -- think of, e.g., trapped ions \cite{Leibfried:2003}, cold atoms \cite{Zenesini:2009}, or color centers in diamonds \cite{Fuchs:2009}. In the presence of a driving field, the system Hamiltonian is time-dependent, and the notion of eigenstates becomes obsolete. However, for time-periodic driving, the \textit{dressed state}, or \textit{Floquet}, picture applies, constituting a framework that is essentially equivalent to the concept of eigenstates of autonomous systems \cite{Shirley:1965,Cohen-Tannoudji:1969,Buchleitner:2002}. In particular, it provides quasi-stationary solutions of the dynamics, i.e., solutions for which the system dynamics repeats itself periodically after each driving cycle. Thus, if one of the Floquet states is strongly entangled over the entire driving period, and if the system is initially prepared in this state, its entanglement is preserved -- in complete analogy to entangled eigenstates of autonomous systems.
Nevertheless, beyond this formal analogy, Floquet states can differ strongly in their properties from the eigenstates of the respective undriven system. This is manifest in micromaser physics \cite{Meschede:1985,Haroche:1992}, rapid adiabatic passage \cite{Bergmann:1998}, the coherent destruction of tunneling \cite{Grossmann:1991}, non-dispersive wave packets \cite{Buchleitner:2002}, scenarios of Anderson / dynamical localization in light-matter interaction \cite{Blumel:1991}, and the asymptotic persistence of quantum coherence under environment coupling \cite{Haroche:1992,Wellens:2000,Buchleitner:2002a}. 

In this paper, we show how maximally entangled Floquet states of interacting two-level systems can be generated by controlling the parameters of an external, time-periodic driving field. Taking a perturbative approach in the interaction strength between the qubits, we connect their occurrence to avoided crossings of many-particle states in the quasi-energy spectrum. This is conceptually closely related to controlled $N$-particle interactions that have recently been demonstrated in a cold Rydberg gas \cite{Gurian:2012}, only that here we assume alternating instead of static control fields, what enormously widens the versatility of the control strategy. For autonomous quantum systems, the connection between avoided crossings and entanglement has been discussed earlier, as, e.g., in spins chains \cite{Karthik:2007,Bruss:2005}. Our generalization to periodically driven system opens up the possibility of creating strong, stationary entanglement in multi-partite quantum systems, by simply tuning the parameters of an external driving field.


%

%% file: Theory.tex

\section{Theoretical framework}

\subsection{Floquet theory}\label{ssec:Floqtheory}

We start with a recollection of Floquet theory, on which we rely throughout this work, and which is the semiclassical \footnote{Floquet theory is semiclassical in the sense that the driving field is not treated in a quantized fashion, but as time-dependent term in the Hamiltonian.} variant of dressed state theory \cite{Shirley:1965,Cohen-Tannoudji:1969}. It is founded on the theorem that, given a Hamiltonian $H(t)=H(t+T)$ with period $T$, every solution of the Schr\"odinger equation
\begin{equation}\label{eq:Schroedinger_eq}
	i  \, \partial_t \ket{\Psi(t)} = H(t) \ket{\Psi(t)} \qquad (\hbar \equiv 1)
\end{equation}
can be written as a superposition
\begin{equation}\label{eq:Floq_state_def}
	\ket{\Psi(t)} = \sum_i a_i\, e^{-i \varepsilon_i t} \ket{\Phi_i(t)}
\end{equation}
of mutually orthogonal, $T$-periodic \textit{Floquet states} $\ket{\Phi_i(t)}=\ket{\Phi_i(t+T)}$. The \textit{quasi-energies} $\varepsilon_i$ appearing in the phase factor are real numbers, and the time-independent weighting factors of the superposition are $a_i=\braket{\Phi_i(0)|\Psi(0)}$. Hence, once the Floquet states and quasi-energies of $H(t)$ are known, the time evolution of the system is  available for arbitrary times and arbitrary initial states.

The Floquet picture can be regarded as the generalization of the concept of eigenstates to periodically driven systems: Consider, for the moment, an autonomous Hamiltonian $H_0$, plus a weak periodic perturbation $V(t)=V(t+T)$, such that $H(t) = H_0+\lambda V(t)$. In the limit $\lambda \rightarrow 0$, a possible choice of Floquet states and quasi-energies are, respectively, the eigenvectors $\ket{\Phi_i}$ and eigenvalues $E_i$ of $H_0$, which can be found by solving the eigenvalue problem $H_0 \ket{\Phi_i} = E_i \ket{\Phi_i}$. In this case, Eq.~\eqref{eq:Floq_state_def} is simply the well-known time evolution of eigenstates, reflecting the fact that an autonomous system will not leave an initially prepared eigenstate. For finite $\lambda$, the Floquet states of $H(t)$ become time-dependent. As long as the driving $V(t)$ varies only slowly in time, the system dynamics will be adiabatic; i.e., the Floquet states are the instantaneous eigenstates of $H(t)$, given by $H(t) \ket{\Phi_i(t)} = E_i(t) \ket{\Phi_i(t)}$. However, for fast (and/or strong) driving, the adiabaticity assumption is not justified (cf. Refs.~\cite{Breuer:1989,Tong:2007} for a detailed discussion), and Floquet states no longer coincide with the instantaneous eigenstates of $H(t)$.
Nevertheless, Eq.~\eqref{eq:Floq_state_def} still implies that after initially preparing a Floquet state $\ket{\Phi_k(t)}$ (i.e., $a_i=\delta_{ik}$), the system remains in this state for arbitrarily long times, and only gains a dynamical phase $e^{-i \varepsilon_k t}$. The expectation values of any (time-independent) observable then exhibits at most $T$-periodic time-dependence.

As in the static case, Floquet states can be transferred into each other via frequency-matched probe fields \cite{Breuer:1988}, or by adiabatic passage \cite{Buchleitner:2002}.
An important difference to the concept of eigenstates of autonomous systems is, however, that Floquet states and quasi-energies are never unique: Introducing the driving frequency $\omega=\frac{2\pi}{T}$, one finds that with every Floquet state $\ket{\Phi_i(t)}$, $e^{-i k \omega t}\ket{\Phi_i(t)}$ is an equivalent Floquet state, for arbitrary integer $k$. This is because the latter expression retains $T$-periodicity and, after inserting it into Eq.~\eqref{eq:Floq_state_def} and shifting the quasi-energy $\varepsilon_i$ by $k \omega$, leads to the same solution $\ket{\Psi(t)}$ of the Schr\"odinger equation \cite{Friedrich:2006}. (More formally, this equivalence relation defines a rest class structure in the set of Floquet states \cite{Breuer:1989a}.) As a consequence, the spectrum of quasi-energies is identical in every interval $[k \omega, k \omega + \omega)$. By restricting quasi-energies to a single such \textit{Floquet zone}, one gets rid of the ambiguity and is left with complete and mutually orthogonal set of $d$ Floquet states, $d$ being the dimension of the Hilbert space that $H(t)$ is acting upon.

\subsection{Floquet state entanglement}

In this paper, we study the entanglement of Floquet states of a periodically driven, closed quantum system, with the following motivation: If the system is initially prepared in one of its Floquet states $\ket{\Phi_i(0)}$, the solution of the Schr\"odinger equation is, according to Eq.~\eqref{eq:Floq_state_def}, $\ket{\Psi(t)} = e^{-i \varepsilon_i t} \ket{\Phi_i(t)}$. Since a global phase factor is irrelevant to entanglement \footnote{This argument also guarantees that all Floquet states of the same rest class have identical entanglement properties, since they differ only by a phase factor $e^{-i k\omega t}$. Hence, entanglement properties of Floquet states do not depend on the Floquet zone of the spectrum one is looking at.}, the system remains entangled as long as $\ket{\Phi_i(t)}$ is. At best, $\ket{\Phi_i(t)}$ is maximally entangled over the entire driving period $t\in [0,T)$.
The system then remains maximally entangled for, in principle, arbitrarily long times. For this reason, it is desirable to understand under which conditions Floquet states are strongly, maybe even maximally, entangled.

Of course, under realistic conditions, decoherence will affect the system after a certain time and lead to a deviation from the perfectly coherent scenario described by a pure quantum state, that we study in this work. Nevertheless, even in presence of weak environment coupling, Floquet states are typically the most robust states of a periodically driven quantum system \cite{Blumel:1991}, and therefore highly entangled Floquet states are advantageous also in presence of decoherence \cite{Timoney:2011}.

To quantify the entanglement $\overline{\mathcal{E}}_i$ of a Floquet state $\ket{\Phi_i(t)}$, we consider the time average of a predefined entanglement measure \cite{Mintert:2005} $\mathcal{E}$ over one period  $T$:
\begin{equation}\label{eq:def_mean_ent}
\overline{\mathcal{E}}_i\equiv \frac{1}{T} \int_0^{T} \mathcal{E}(\ket{\Phi_i(t)})\, \mathrm{d}t.
\end{equation}
(This is reasonable, since $\mathcal{E}(\ket{\Phi_i(t)})$ is $T$-periodic quantity.)
The particular choice of $\mathcal{E}$ will depend on the number of the two-levels under investigation in the subsequent sections. Irrespectively of this choice, however, one always has maximal Floquet state entanglement $\overline{\mathcal{E}}_i=1$ if and only if $\ket{\Phi_i(t)}$ is maximally entangled \textit{for all times} $t\in [0,T)$, since every (normalized) entanglement measure is a non-negative function that vanishes for separable states and takes its maximum value $\mathcal{E}=1$ for maximally entangled states. Vice versa, Floquet state entanglement vanishes if and only if $\ket{\Phi_i(t)}$ is separable for all $t\in [0,T)$.


\subsection{Mathematical structure of the Floquet problem}\label{sec:Floquet_details}
Before we investigate Floquet state entanglement in detail, we elaborate on the mathematical structure of the Floquet problem, which consists in finding the Floquet states $\ket{\Phi_i(t)}$ and respective quasi-energies $\varepsilon_i$, for a given time-periodic Hamiltonian $H(t)$. This discussion will help us to analyze the phenomena encountered later. 

We start by inserting \eqref{eq:Floq_state_def} into \eqref{eq:Schroedinger_eq}, what leads to
\begin{equation}\label{eq:Floq_eigv_problem}
	[H(t) - i \partial_t] \ket{\Phi_i(t)} = \varepsilon_i \ket{\Phi_i(t)}.
\end{equation}
This equation is reminiscent of the eigenvalue problem $H\ket{\Phi_i}=E_i\ket{\Phi_i}$ for autonomous quantum systems, and implies that Floquet states and quasi-energies are eigenvectors and eigenvalues of the \textit{Floquet Hamiltonian} $\mathbf{H}_F\equiv H(t) - i \partial_t$. Differently to the static case, however, $\mathbf{H}_F$ acts on the extended \textit{Floquet Hilbert space} $\mathcal{H}_F\equiv \mathcal{H}\otimes \mathrm{L}^2([0,T))$ of all $T$-periodic orbits in the original Hilbert space $\mathcal{H}$. Since the space $\mathrm{L}^2([0,T))$ of all $T$-periodic, square-integrable functions is isomorphic to $\ell^2$ \cite{Katznelson:2004}, it is spanned by a discrete basis set, e.g., $\{e^{-i k \omega t}, k\in\mathbb{Z}\}$. Hence, every periodic orbit $\ket{\Phi_i(t)}$ can be expanded in a discrete Fourier series
\begin{equation}\label{eq:Floqstat_timedomain}
	\ket{\Phi_i(t)} =  \sum_{k\in\mathbb{Z}} \ket{\tilde\Phi_i(k)}e^{-i k \omega t}
\end{equation}
with
\begin{equation}
 \ket{\tilde\Phi_i(k)} = \frac{1}{T} \int_0^T dt \, \ket{\Phi_i(t)} e^{i k \omega t}.
\end{equation}
The periodic orbit is then represented by a ``double ket'' in $\mathcal{H}_F$,
\begin{equation}\label{eq:Floqstate}
	\ket{\ket{\Phi_i}} \equiv \sum_{k\in\mathbb{Z}} \ket{\tilde\Phi_i(k)}\otimes\ket{k},
\end{equation}
where $\ket{k}$ denotes the Fourier basis functions $e^{-i k \omega t}$ of $\mathrm{L}^2([0,T))$.
After expressing the Floquet Hamiltonian $\mathbf{H}_F$ in the Fourier basis as well,
\begin{equation}\label{eq:Floq_Hamiltonian}
	\mathbf{H}_F = \sum_{k,l \in \mathbb{Z}} [\tilde H_{l-k} + \delta_{kl} k \omega] \otimes \ket{k}\bra{l},
\end{equation}
with
\begin{equation}\label{eq:Hamiltonian_Fourier}
	 \tilde H_{l-k} = \frac{1}{T} \int_0^T dt \, H(t) e^{i (l-k)\omega t} = \tilde H^\dagger_{k-l},
\end{equation}
Eq.~\eqref{eq:Floq_eigv_problem} reads
\begin{equation}\label{eq:Floq_eigv_problem2}
	 \mathbf{H}_F \ket{\ket{\Phi_i}} = \varepsilon_i \ket{\ket{\Phi_i}}.
\end{equation}
Writing out the Fourier indices $k$ and $l$ in matrix form instead, we see that Floquet states and eigenvalues of $H(t)$ can be found by diagonalizing the matrix
\begin{eqnarray}\label{eq:Floq_Hamiltonian2}
	\mathbf{H}_F=\left( \begin{matrix}
	  \ddots & & \vdots & & \\
	   & \tilde H_0 + \omega & \tilde H_1 & \tilde H_2 &  \\
           \hdots& \tilde H^\dagger_1 & \tilde H_0  & \tilde H_1&  \hdots \\
	  & \tilde H^\dagger_2 & \tilde H^\dagger_1 &  \tilde H_0 - \omega &   \\
	  & & \vdots & & \ddots \\
	\end{matrix}\right).
\end{eqnarray}
This representation of the Floquet problem has a few interesting implications:
\begin{enumerate}

\item A major advantage of rephrasing the Schr\"odinger equation as eigenvalue problem in Eq.~\eqref{eq:Floq_eigv_problem2} is that all the concepts developed for autonomous quantum systems can be imported. In particular, time-\textit{in}dependent perturbation theory can be applied to Eq.~\eqref{eq:Floq_eigv_problem2} \cite{Shirley:1965}. For example, assume that the Floquet states $\ket{\ket{\Phi_i}}$ of a certain $T$-periodic $H(t)$ are known. The impact of a small perturbation $V(t)$ (of the same periodicity) is then determined, to first order, by matrix elements
\begin{equation}
	 c_{ij}=\braket{\braket{\Phi_i || \mathbf{H}^p_1 || \Phi_j}} = \sum_{k,l} \braket{\tilde\Phi_i(k) | \tilde V_{l-k} | \tilde\Phi_j(l)}
\end{equation}
in Floquet Hilbert space. In the time domain, this reads
\begin{equation}\label{eq:Cmatrixel_theory}
	 c_{ij}= \frac{1}{T}\int_{0}^{T} dt \braket{\Phi_i(t) | V(t) | \Phi_j(t)}.
\end{equation}
In particular, if two unperturbed quasi-energies $\varepsilon_i$ and $\varepsilon_j$ of $H(t)$ cross (as some system parameter is varied), $V(t)$ can lift this degeneracy and lead to an avoided crossing with minimal level separation $2|c_{ij}|$. 

\item The diagonal of \eqref{eq:Floq_Hamiltonian2} contains the static part $\tilde H_0$ of the system Hamiltonian, while the off-diagonal entries $\tilde H_j$ reflect driving with frequency $j \omega$. E.g., in case of an autonomous Hamiltonian $H$ with eigenstates $\ket{\Phi_i}$ and eigenvalues $E_i$, we have $\tilde H_0=H$ and all other $\tilde H_j=0$. Then, $\ket{\Phi_i} \otimes \ket{0}$ is an eigenstate of $\mathbf{H}_F$ with quasi-energy $E_i$; but all $\ket{\Phi_i} \otimes \ket{k}$ are eigenstates as well, with quasi-energies $E_i + k \omega$. This reflects precisely the previously discussed rest class structure of Floquet states.

\item Despite the fact that the driving field is not quantized in the Floquet picture, but is rather represented by a time-dependent classical field in the Hamiltonian, the Fourier index $k$ in Eqs.~\eqref{eq:Floqstat_timedomain}-\eqref{eq:Hamiltonian_Fourier} can be interpreted as the number of quanta $\omega$ in the driving field. In fact, an exact correspondence between the Floquet Hamiltonian in the Fourier basis and the ``dressed'' Hamiltonian of the quantized case can be established for large occupation numbers of the quantized mode \cite{Shirley:1965, Cohen-Tannoudji:1973}.

\item Being interested in $N$ two-level systems in this paper, the extended Floquet Hilbert space is $\mathcal{H}_F=[\mathbb{C}^2 ]^{\otimes N} \otimes \mathrm{L}^2([0,T))$ throughout the sequel of this paper. One can formally interpret this as the Hilbert space of an $(N+1)$-partite quantum system (the $(N+1)$-th particle having not just two, but countably infinitely many levels), with dynamics generated by the \textit{autonomous} Hamiltonian $\mathbf{H}_F$.

\item The number of non-vanishing Fourier components $\ket{\tilde\Phi_i(k)}$ is determined by the structure of the driving terms $\tilde H_j$ in \eqref{eq:Floq_Hamiltonian2} (see, e.g., Ref.~\cite{Buchleitner:2002}), and decays with large $|k|$.
Hence, the Fourier index $k$ can effectively be confined to finite intervals $-M<k\le M$. Then, $\mathbf{H}_F$ becomes a finite $(2M\, 2^N) \times (2M\, 2^N)$ matrix, which can be diagonalized by standard numerical methods. In \ref{sec:AppendixNoFouriercomp}, we derive the rule of thumb $M \approx 2 F/\omega$ for a qubit that is driven by a monochromatic field of amplitude $F$ and frequency $\omega$.
\end{enumerate}

%% file: 2Qubits.tex

\section{Two weakly coupled qubits under external driving}\label{sec:N2}

As starting point of our investigation, we consider two two-level systems (qubits) that are driven by an external field with periodic amplitude $f(t)=f(t+2\pi / \omega)$, and coupled by a $\sigma_+\sigma_-$ ``excitation exchange'' interaction. The Hamiltonian reads
\begin{equation}\label{eq:Hamiltonian2}
	H(t) = \sum_{n=1}^{2} \left( \frac{{\omega_0}}{2}  \sigma_{z}^{(n)} + f(t) \sigma_x^{(n)} \right)  + H_\textrm{qq},
\end{equation}
\begin{equation}\label{eq:Hqq2}
	H_\textrm{qq} = C \left( \sigma_{+}^{(1)} \sigma_{-}^{(2)} + \sigma_{-}^{(1)} \sigma_{+}^{(2)}\right),
\end{equation}
with Pauli operators $\sigma^{(n)}$ acting on the $n$-th qubit, ${\omega_0}$ the single qubit energy splitting, and $C$ the qubit-qubit interaction strength. The term $f(t) \sigma_x^{(n)}$ describes a coherent, classical electro-magnetic driving field. We assume it to be identical for both qubits, so that no individual addressability of the qubits is required here.

Hamiltonian \eqref{eq:Hamiltonian2} is encountered in various physical scenarios, prominent examples of which are listed in Table \ref{tab:physicalrealizations}. In case of trapped ions \cite{Blatt:2008} and superconducting qubits \cite{Steffen:2006}, entanglement is routinely being measured in experiments, e.g., by state tomography. The entanglement of Floquet states, as investigated presently, can be demonstrated experimentally in the very same fashion.
Note that in some of the scenarios of Table~\ref{tab:physicalrealizations}, the qubit-qubit interaction $H_{qq}$ is not described by a $\sigma_+\sigma_-$ excitation exchange mechanism. However, as long as qubits do interact pairwise in some form, the details of $H_{qq}$ are not crucial for the phenomena discussed in the following, as we will argue in Section \ref{sec:interactionvariation}. 


\begin{table}
\centering
\begin{tabularx}{\linewidth}{p{.135\textwidth}||X|X|X|X}
 & supercon\-duc\-ting qubits \newline \cite{Devoret:,McDermott:2005,Niemczyk:2010a}	& trapped ions \newline \cite{Ospelkaus:2011,Khromova:2011,Timoney:2011}  & Color centers in diamond \newline \cite{Fuchs:2009,Neumann:2010} & Rydberg atoms \cite{Gallagher:2008mz,Urban:2009,Wilk:2010} \\ \hline\hline  
 driving source 		& microwave & laser or \newline microwave & microwave & laser  \\ \hline  
 qubit-qubit interaction 	& inductively /\newline capacitively, or via cavity & phonon-mediated & dipole-dipole & dipole-dipole 	\\ \hline %
 $\omega_0 / 2\pi$ 		& 5 - 20 GHz		&  1 - 10 GHz		& 0.5 - 3 GHz 		& 2.4$\cdot 10^{14}$ Hz		\\ \hline	
 $\omega / 2\pi$ 		& MHz - GHz		&  MHz - GHz		& MHz - GHz		& $10^{14}$ - $10^{15}$ Hz	\\ \hline	 
 $F / 2\pi$ 			& 50 - 500 MHz		&  30 MHz			& 30 - 400 MHz		& 0.5 - 5 MHz				\\ \hline  
 $C / 2\pi$ 			& 80 MHz			&  30 Hz  			& 40 kHz			& 50 MHz 	
\end{tabularx}
\caption{Energy scales for various physical realizations of the Hamiltonian \eqref{eq:Hamiltonian2}. All values are exemplary and meant to indicate orders of magnitude. (In case of trapped ions, numbers refer to microwave driving.) Subject of this article is the optimal choice of the amplitude $F$ and of the frequency $\omega$ of the driving field $f(t)$, in order to maximize Floquet state entanglement. }
\label{tab:physicalrealizations}
\end{table}

Since the Hamiltonian \eqref{eq:Hamiltonian2} does not distinguish one or the other qubit, its Floquet states can be chosen as either symmetric or antisymmetric under exchange of the qubits. The antisymmetric subspace of two qubits, however, consists of the singlet $\frac{1}{\sqrt{2}}\left(\ket{\uparrow\downarrow}-\ket{\downarrow\uparrow} \right)$ only. (We adopt the spin-1/2 notation $\ket{\uparrow}$ and $\ket{\downarrow}$ for the qubit basis throughout this paper.) Therefore, this maximally entangled state is always one of the four Floquet states of the Hamiltonian \eqref{eq:Hamiltonian2}, independently of the system parameters. Consequently, we will restrict our discussion to the remaining three Floquet states in the symmetric subspace, which is spanned by $\{\ket{\uparrow\uparrow},\frac{1}{\sqrt{2}}\left(\ket{\uparrow\downarrow}+\ket{\downarrow\uparrow} \right),\ket{\downarrow\downarrow}\}$.

\subsection{Monochromatic driving}
\begin{figure*}[tb]
\begin{tabular}{ c | c  c  c }
	 &  $\overline{\mathcal{E}}_{1}$						& $\overline{\mathcal{E}}_{2}$						    & $\overline{\mathcal{E}}_{3}$ \\ \hline
	\begin{sideways}\parbox{.25\textwidth}{$C =  0.02\,\omega$}\end{sideways}	& \includegraphics[width=.32\textwidth]{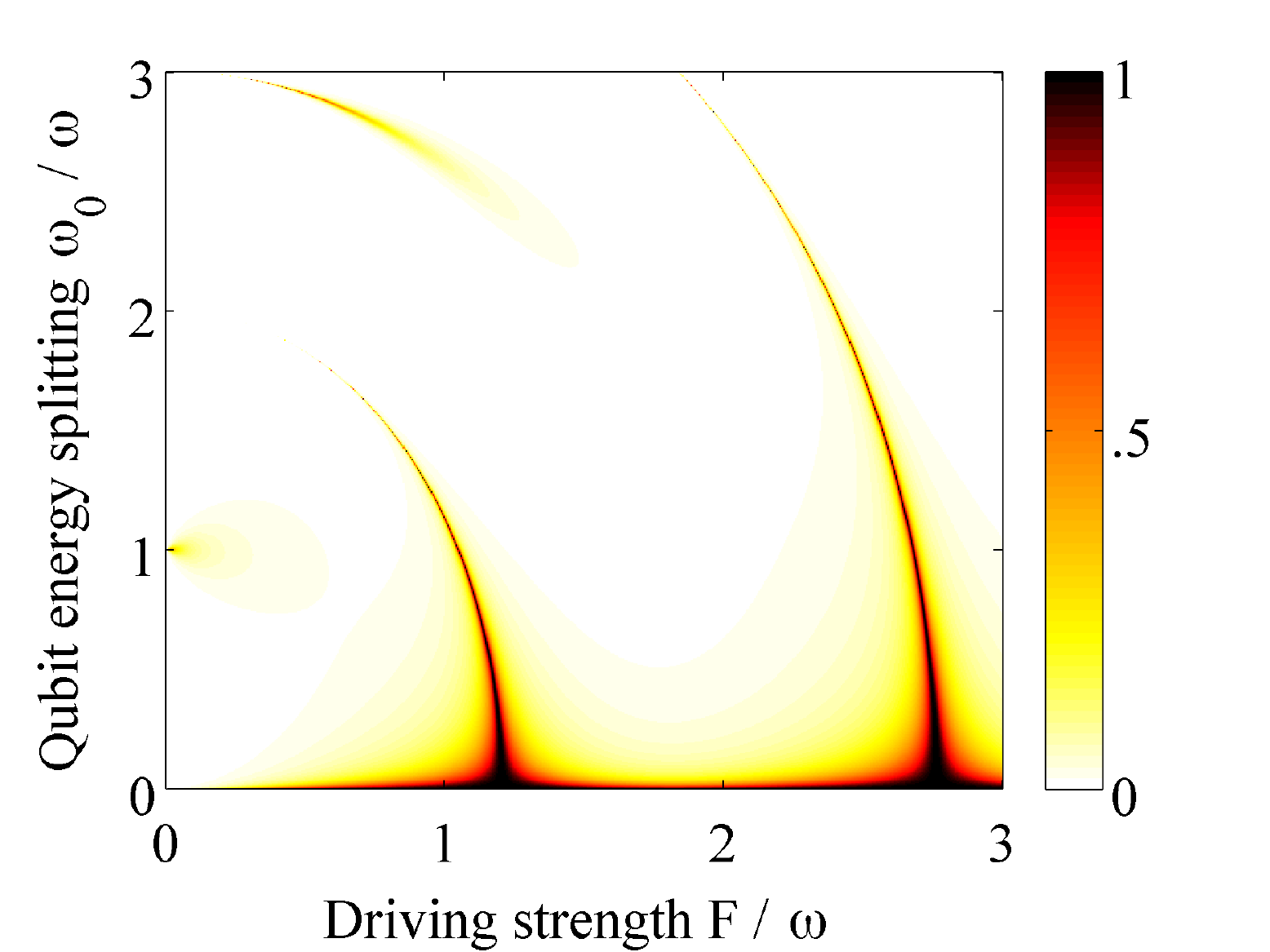} & \includegraphics[width=.32\textwidth]{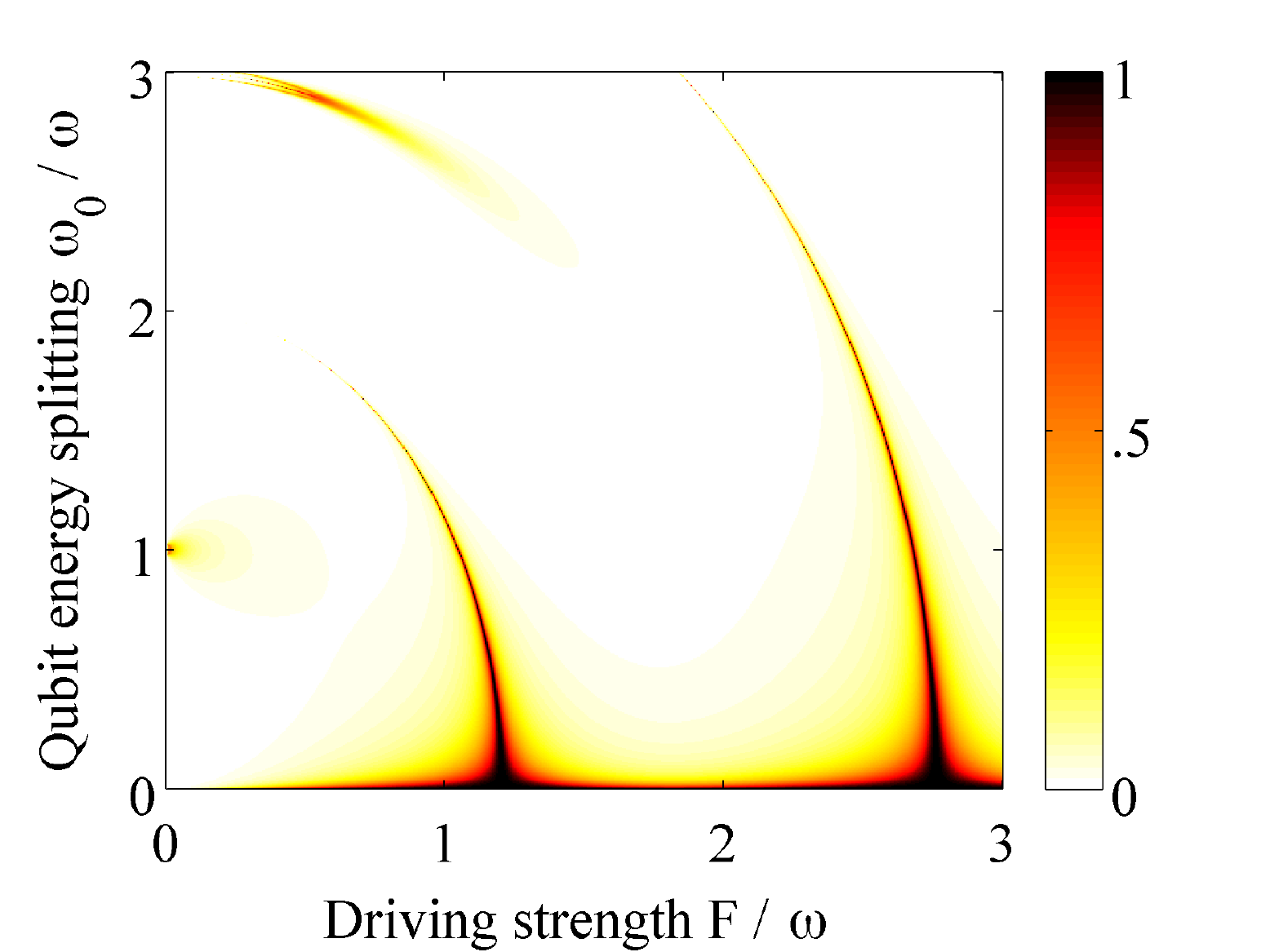} & \includegraphics[width=.32\textwidth]{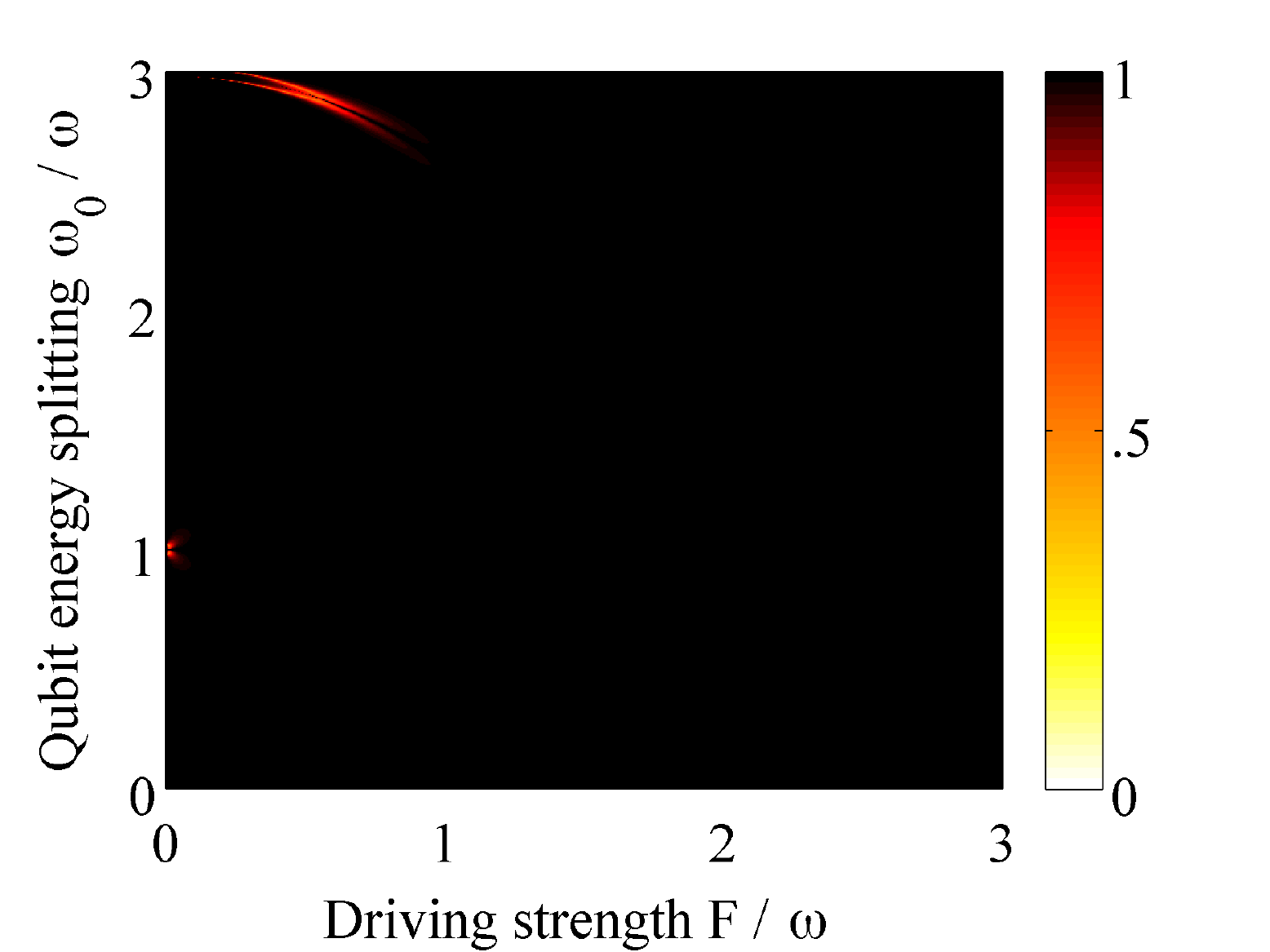}\\
	\begin{sideways}\parbox{.25\textwidth}{$C =  0.2\,\omega$}\end{sideways} & \includegraphics[width=.32\textwidth]{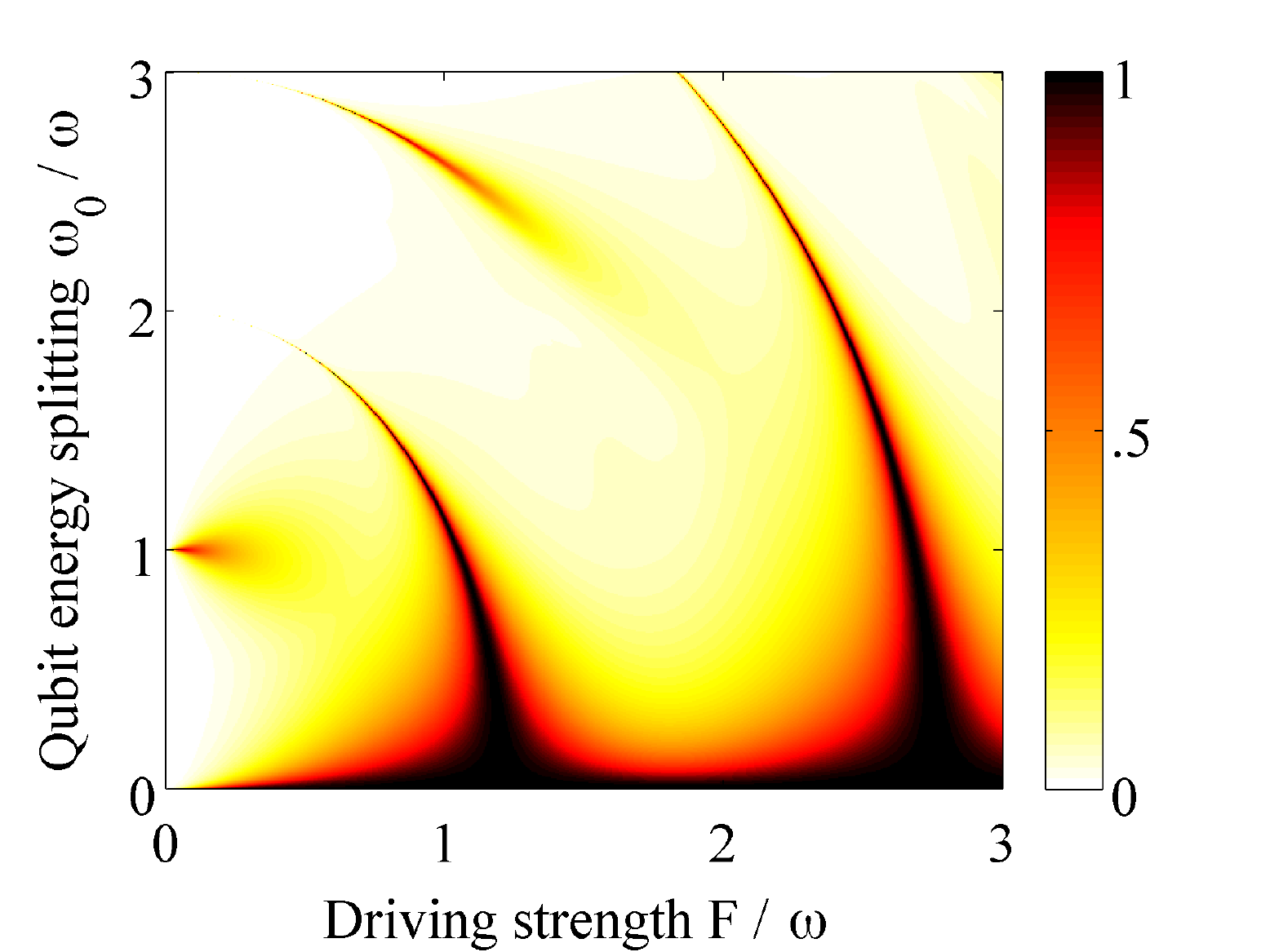} & \includegraphics[width=.32\textwidth]{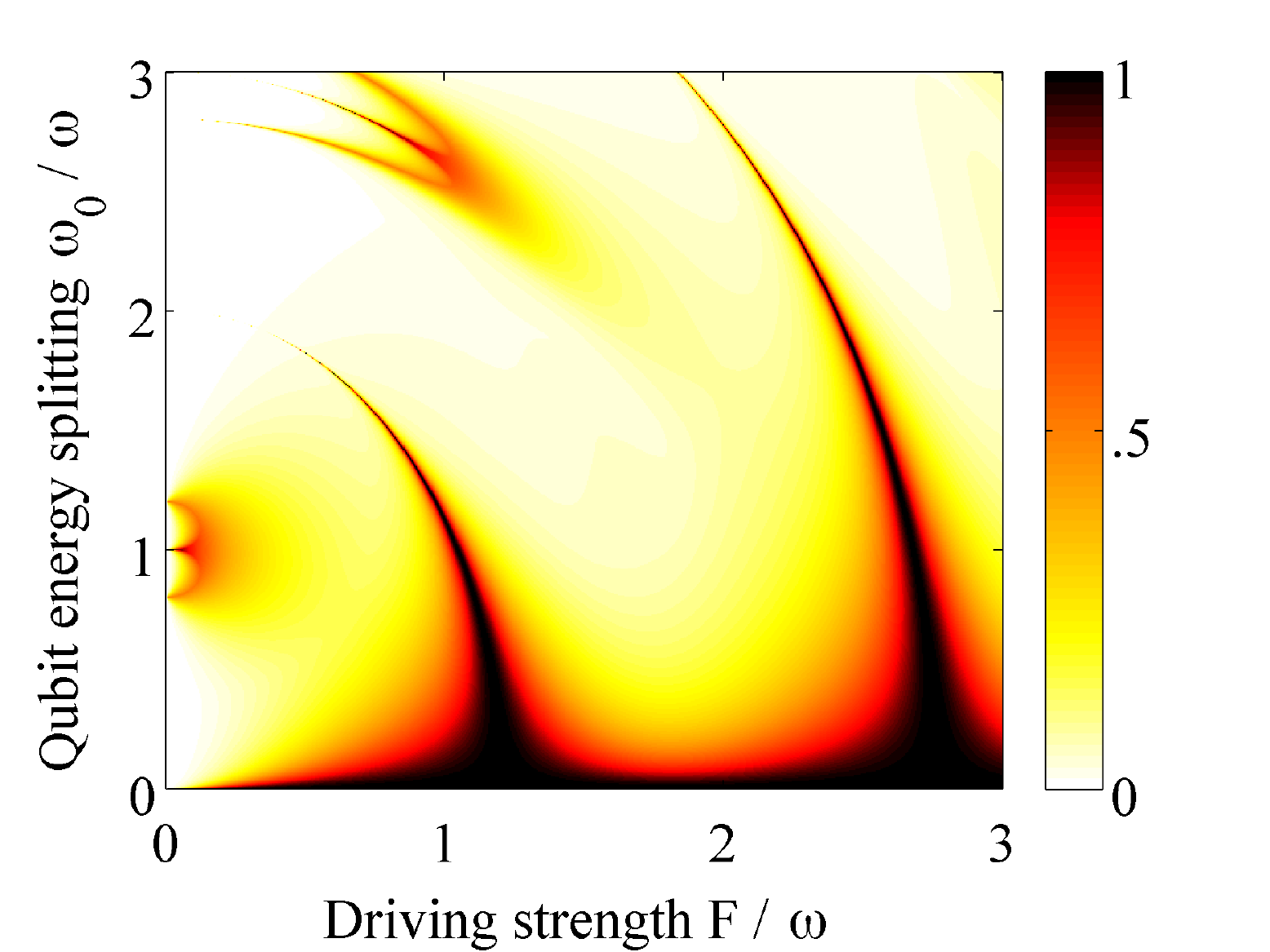} & \includegraphics[width=.32\textwidth]{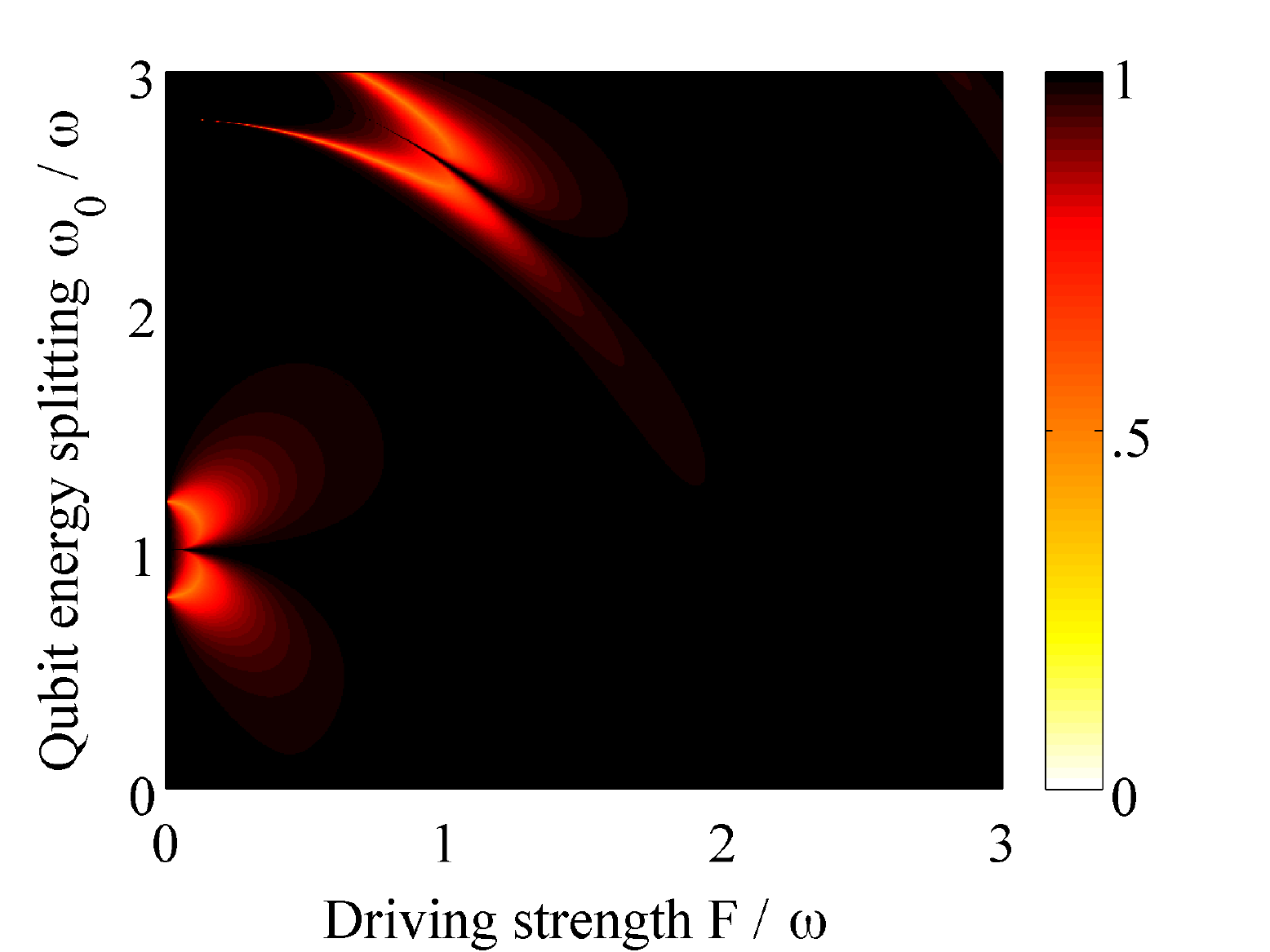}\\
	\begin{sideways}\parbox{.25\textwidth}{$C =  2\,\omega$}\end{sideways} & \includegraphics[width=.32\textwidth]{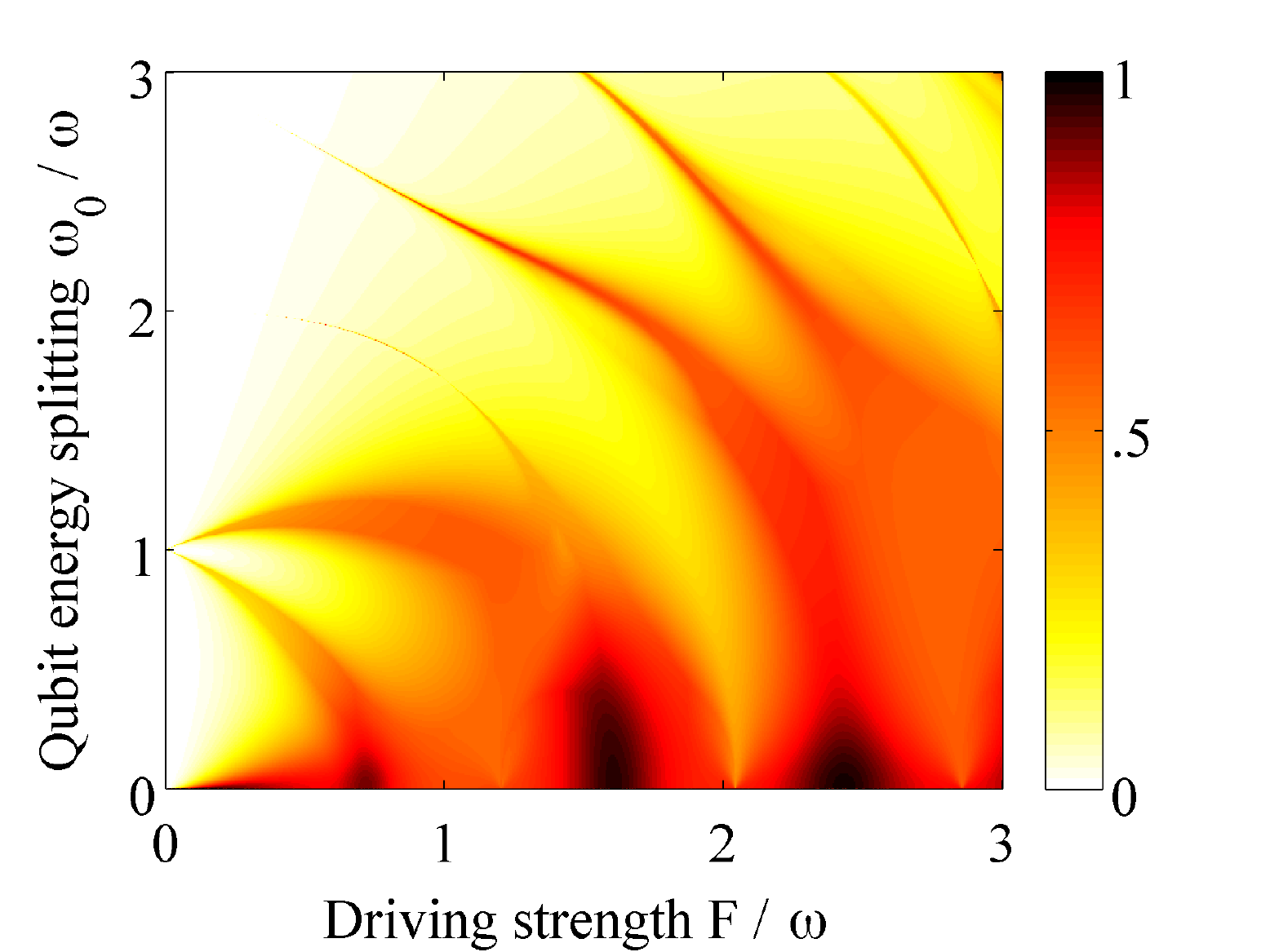} & \includegraphics[width=.32\textwidth]{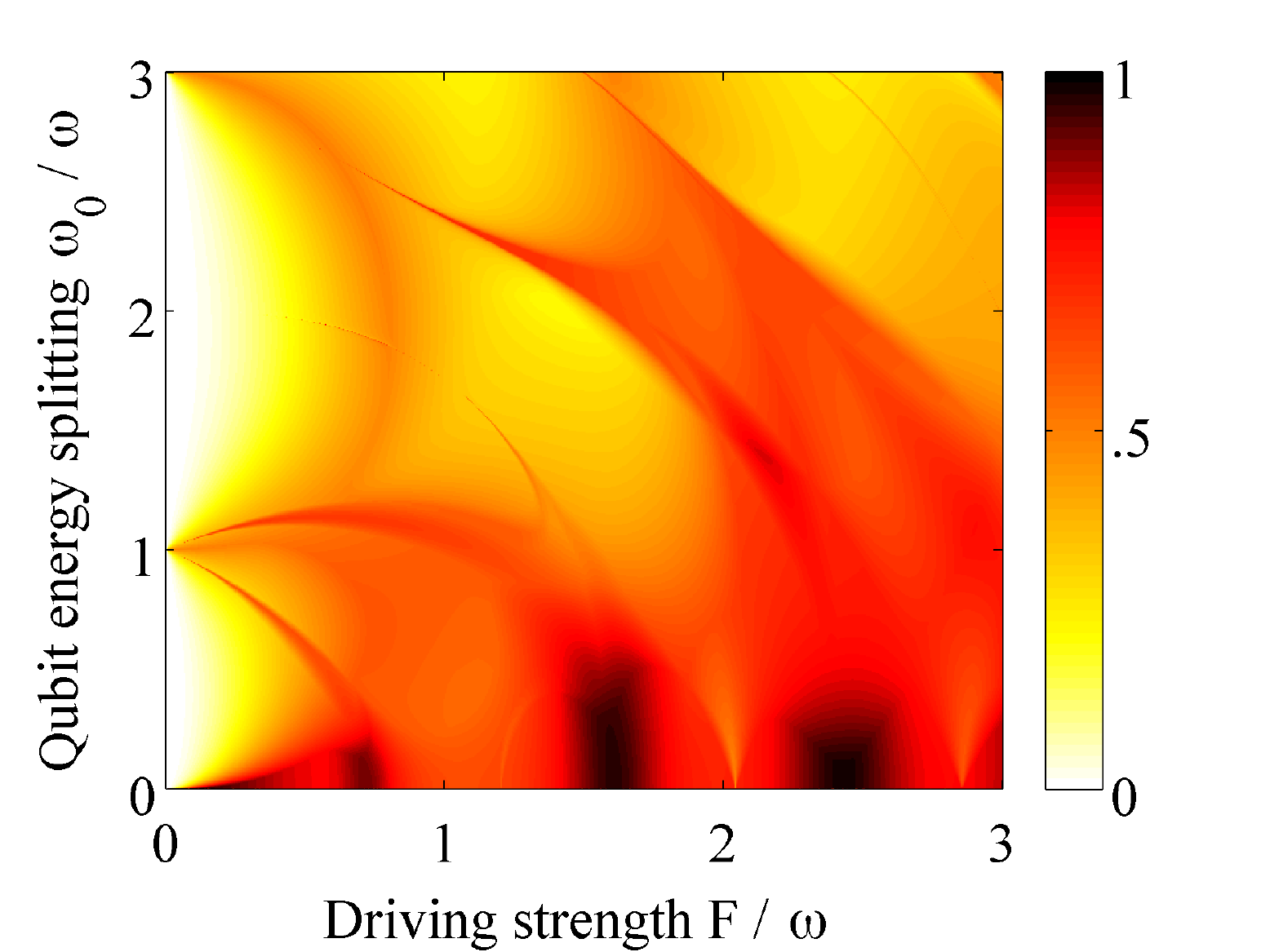} & \includegraphics[width=.32\textwidth]{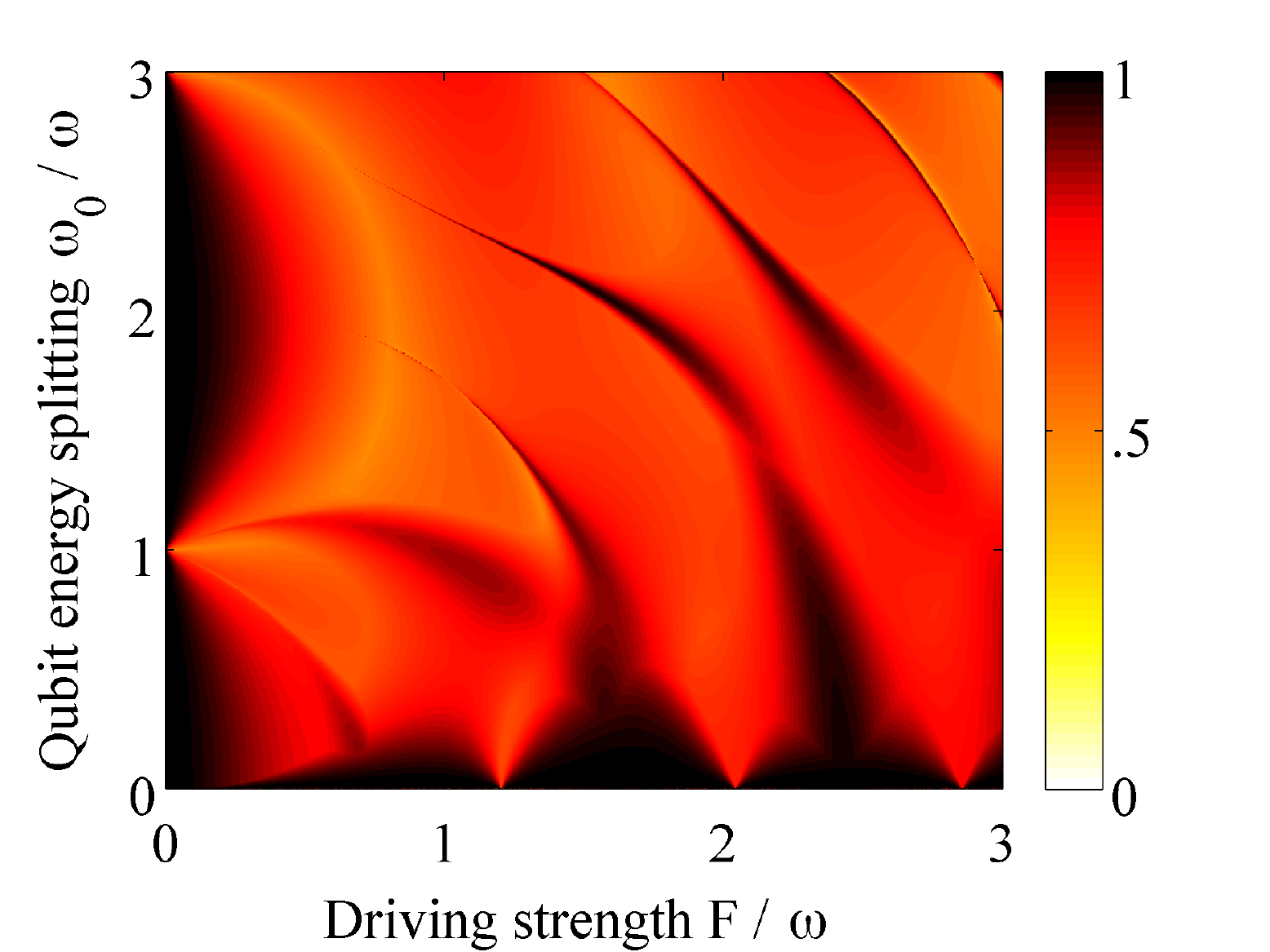}\\
\end{tabular}
\caption{Floquet state entanglement $\overline{\mathcal{E}}_i$ of two coupled qubits under monochromatic driving. Each plot shows the parameter plane spanned by driving amplitude $F$ and qubit energy splitting $\omega_0$, both measured in units of the driving frequency ${\omega}$. The entanglement is visualized by a color code ranging from white (separable) to black (maximally entangled). The three columns correspond to the three Floquet states of the symmetric subspace. (The remaining forth Floquet state is anti-symmetric and \textit{always} maximally entangled.) In the upper row, where the qubit-qubit coupling $C=0.02\,\omega$ is weak, the first two Floquet states have close-to-zero entanglement in large parts of the parameter space, and only along sharp ridges do entanglement ``resonances'' emerge. As $C$ increases in the middle and bottom row, the resonances become broader and overlap.
}\label{fig:N2_2Dplots}
\end{figure*}

An easily implementable driving scheme is monochromatic driving with frequency $\omega$ and amplitude $F$, i.e., $f(t) = F \, \cos(\omega t)$ in Eq. \eqref{eq:Hamiltonian2}.

Figure~\ref{fig:N2_2Dplots} shows the entanglement $\mathcal{E}_i$ (indicated by the color code) of the three symmetric Floquet states, determined by numerically diagonalizing the Floquet problem~\eqref{eq:Floq_Hamiltonian2}, for different values of the system parameters. As entanglement measure $\mathcal{E}$ in definition \eqref{eq:def_mean_ent}, we choose the concurrence \cite{Hill:1997}
\begin{equation}
	\mathcal{E}(\ket{\psi})=| \braket{\psi^* | \sigma_y \otimes \sigma_y | \psi } |.
\end{equation}
The axes of each plot span the parameter plane of driving strength $F$ and qubit energy $\omega_0$; the three rows show different coupling strengths $C$. All parameters are measured in units to the fourth parameter $\omega$. In this way, we scan the entire parameter range of Hamiltonian \eqref{eq:Hamiltonian2} in Figure~\ref{fig:N2_2Dplots}.

For weak interaction between the qubits, $C=0.02\,\omega$ (top row), the third Floquet state (right column) is almost always maximally entangled. On the contrary, the other two states have vanishing entanglement almost everywhere, except for small values of $\omega_0$ and along narrow ridges in the $F$-$\omega_0$-plane. As $C$ increases (middle and bottom row), these entanglement resonances become broader and begin over overlap.

In order to understand the rich patterns observed on Figure~\ref{fig:N2_2Dplots}, we start with the weakly interacting case ($C\ll f(t),\omega_0$), in which the qubit-qubit interaction is a small perturbation to the non-interacting scenario of $C=0$. The unperturbed Floquet states in the symmetric subspace read
\begin{eqnarray}\label{eq:noninteractingFS}
	\ket{\Phi_1(t)} & = &\ket{\phi_+(t)} \otimes \ket{\phi_+(t)}, \\ \nonumber
	\ket{\Phi_2(t)} & = &\ket{\phi_-(t)} \otimes \ket{\phi_-(t)}, \textrm{ and} \\ \nonumber
	\ket{\Phi_3(t)} & = &\frac{1}{\sqrt{2}} ( \ket{\phi_+(t)} \otimes \ket{\phi_-(t)} + \ket{\phi_-(t)} \otimes \ket{\phi_+(t)} ),
\end{eqnarray}
where $\ket{\phi_\pm(t)}$ denote the two Floquet eigenstates of the single qubit Hamiltonian
\begin{equation}\label{eq:N1Hamiltonian}
	h(t) = \frac{\omega_0}{2} \sigma_z + F \cos(\omega t) \sigma_x.
\end{equation}
$\ket{\Phi_1(t)}$ and $\ket{\Phi_2(t)}$ are separable at all times, hence $\overline{\mathcal{E}}_{1}=\overline{\mathcal{E}}_{2}=0$. $\ket{\Phi_3(t)}$, on the other hand, is local unitarily equivalent to $\frac{1}{\sqrt{2}} ( \ket{\uparrow \downarrow} + \ket{\downarrow \uparrow} )$, and hence maximally entangled at all times, implying $\overline{\mathcal{E}}_{3}=1$. This local unitary equivalence is established by a time-periodic transformation $U_\phi(t) \otimes U_\phi(t)$, with
\begin{equation}\label{eq:lu}
	U_\phi(t) = \ket{\uparrow}\bra{\phi_+(t)} + \ket{\downarrow}\bra{\phi_-(t)};
\end{equation}
the unitarity of $U_\phi(t)$ is guaranteed by the fact that the two single qubit Floquet states $\ket{\phi_\pm(t)}$ are orthonormal at all $t$.
To summarize: At $C=0$, the left and central plot of Figure~\ref{fig:N2_2Dplots} would appear entirely white, and the right one entirely black.

As long as the unperturbed Floquet states have non-degenerate quasi-energies, perturbation theory (applied to the Floquet Hamiltonian $\mathbf{H}_F$, as discussed in Sec.~\ref{sec:Floquet_details}) guarantees that their character is not drastically altered in the presence of a weak qubit-qubit interaction $C$. I.e., their entanglement should approach smoothly the non-interacting values in the limit $C\rightarrow 0$. This reasoning explains why $\mathcal{E}_1$ and $\mathcal{E}_2$ vanish in large parts of the parameter plane in the upper row of Figure~\ref{fig:N2_2Dplots}, and why $\mathcal{E}_3$ is predominantly maximal. Only in the vicinity of degeneracies of Floquet eigenvalues is a deviation from this picture possible. The resonant behavior of entanglement observed in Figure~\ref{fig:N2_2Dplots} must therefore be anchored in near-degeneracies of some quasi-energies.

Thus, in order to understand the shape and position of the entanglement resonances in Figure~\ref{fig:N2_2Dplots}, we have to study the Floquet spectrum of the non-interacting system. To this end, it suffices to know the two quasi-energies $\mu_+$ and $\mu_-$ of the \textit{single} qubit Floquet problem 
\begin{equation}\label{eq:N1FloqProbl}
	[h(t) - i \partial_t ] \ket{\phi_\pm(t)} = \mu_\pm \ket{\phi_\pm(t)}.
\end{equation}
Since the sum of quasi-energies always equals the trace of the time-averaged Hamiltonian \cite{Shirley:1965},
\begin{equation}
	\mu_{+} + \mu_{-}=\frac{\omega}{2\pi}\int_{0}^{2\pi/\omega} \mathrm{Tr}[h(t)] dt \qquad (\textrm{mod } \omega),
\end{equation}
and $\mathrm{Tr}[h(t)]=0$ in our case, the two single qubit quasi-energies $\mu_\pm$ are not independent, but fulfill $\mu_+ = -\mu_-$ (if we work in the central Floquet zone $[-\frac{\omega}{2},\frac{\omega}{2})$). Therefore, we will speak of \textit{the} single qubit quasi-energy $\mu \equiv \mu_+ = -\mu_-$ in the following .

With this, the quasi-energies $\varepsilon_i$ of the unperturbed two-qubit Floquet states $\ket{\Phi_i(t)}$ of Eq.~\eqref{eq:noninteractingFS} read $\varepsilon_1 = 2\mu$, $\varepsilon_2 = -2\mu$, and $\varepsilon_3 = 0$. Thus, the unperturbed levels are degenerate at $\mu=0$. But this is not the only possible degeneracy between unperturbed Floquet states: If, e.g., $\mu = \omega/4$, we have $\varepsilon_1 = \varepsilon_2 + \omega$, and thus $\ket{\Phi_1(t)}$ is degenerate with the shifted Floquet state $e^{i \omega t}\ket{\Phi_2(t)}$, since the latter has quasi-energy $\varepsilon_2 + \omega$. In general, degeneracy occurs whenever $\varepsilon_1 = \varepsilon_2 + n \omega$, i.e., at
\begin{equation}\label{eq:N2degencond}
	\mu = n \frac{\omega}{4} \qquad (n\in \mathbb{N}_0).
\end{equation}
Only in regions of the parameter space where this condition is (approximately) fulfilled will the Floquet states significantly deviate from those of the unperturbed, non-interacting system at $C=0$. 

\begin{figure*}[tb]
\begin{tabular}{l l}
	(a) & (b)\\
	\includegraphics[width=.45\textwidth]{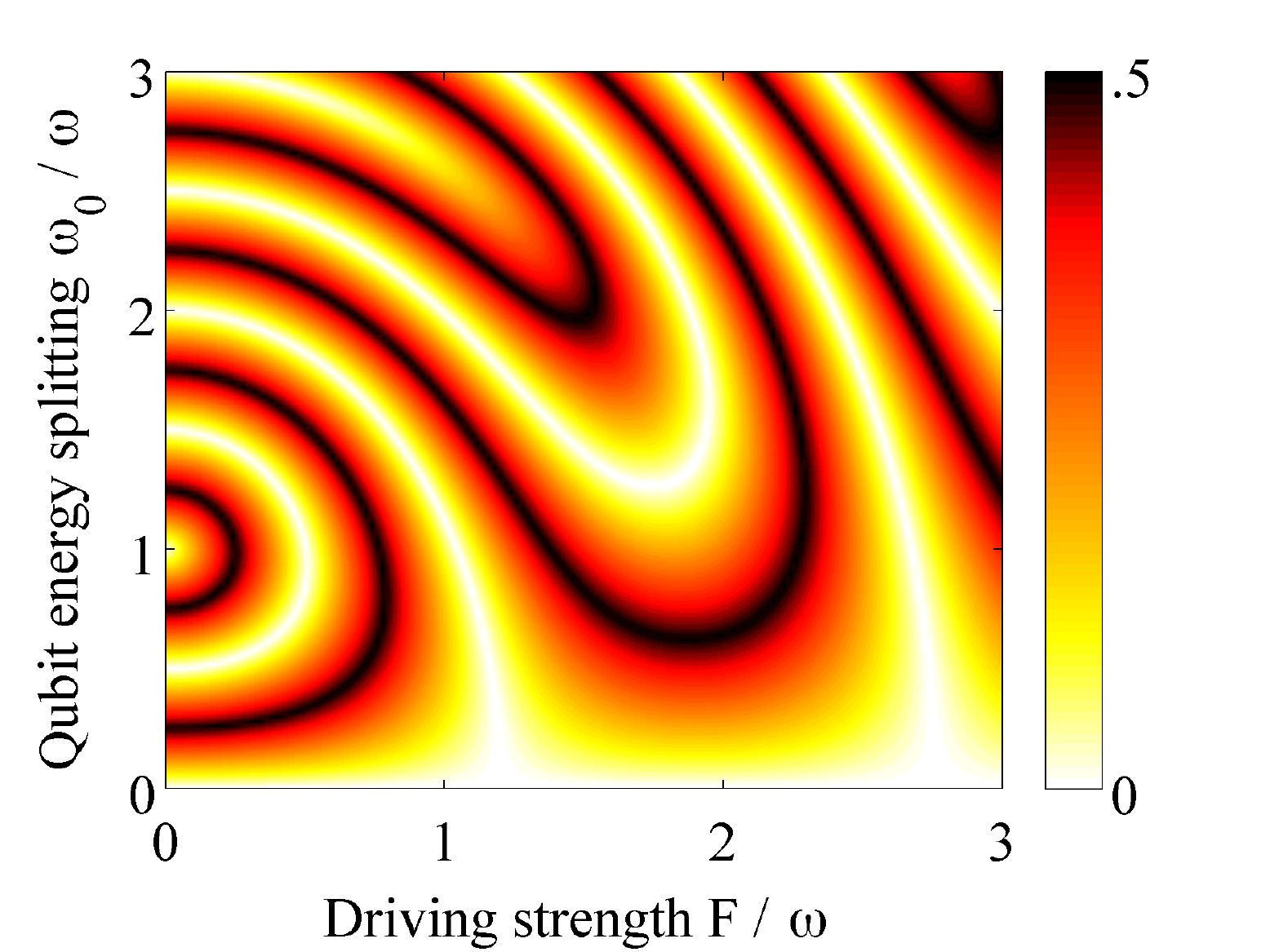} & \includegraphics[width=.45\textwidth]{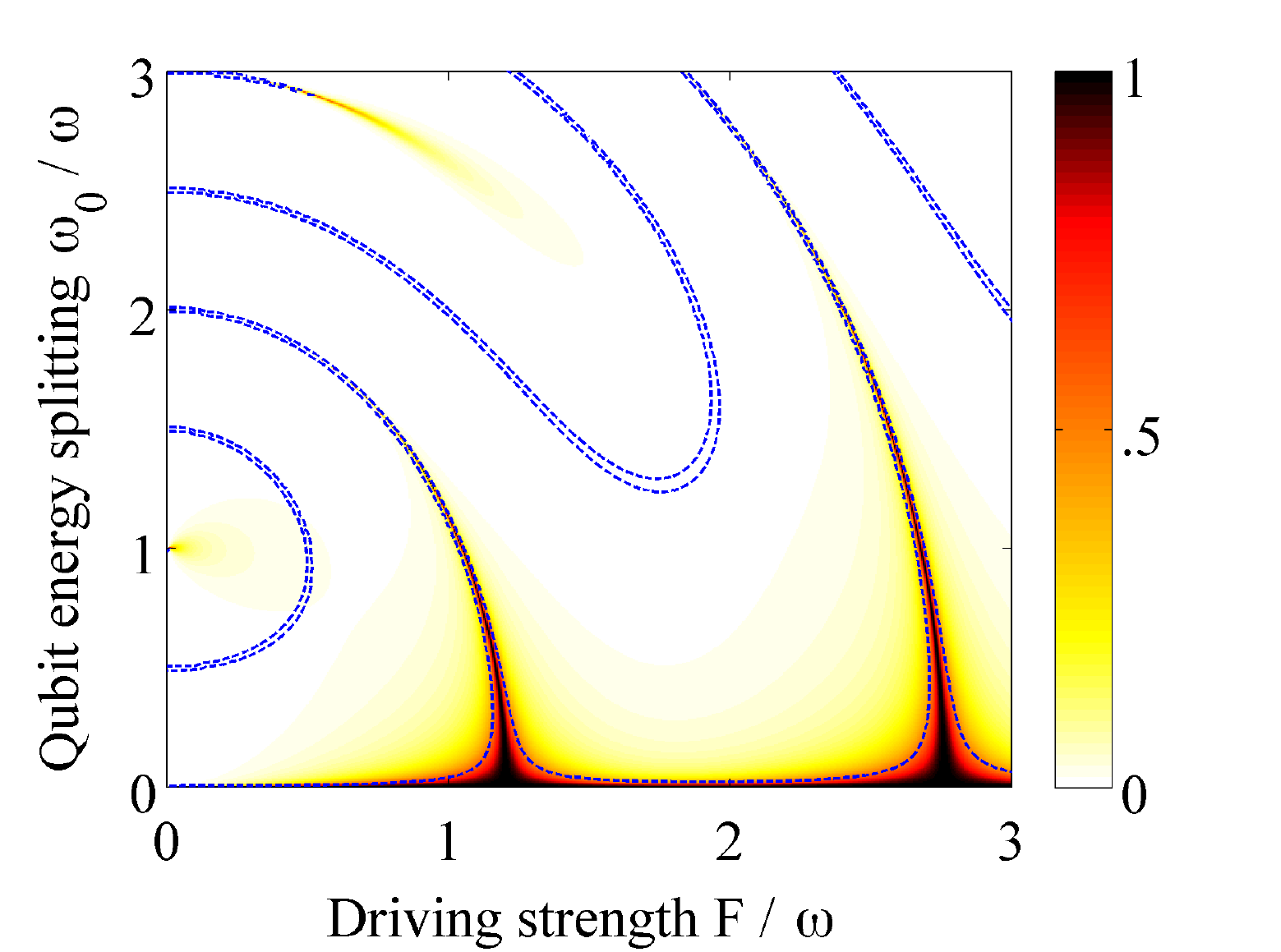}
\end{tabular}
\caption{Detailed analysis of the ``entanglement resonances'' observed in Figure~\ref{fig:N2_2Dplots}, for weak qubit-qubit coupling $C=0.02\,\omega$. (a) Deviation from the degeneracy condition \eqref{eq:N2degencond}, quantified by $\min_n | 4\mu/\omega - n|$, as a function of driving strength $F$ and of the qubit energy splitting $\omega_0$. The color code ranges from white (exact degeneracy) to black (quasi-energies as distant as possible). (b) $\overline{\mathcal{E}}_1$, the entanglement of the least entangled Floquet state in Figure~\ref{fig:N2_2Dplots}, superimposed by contours (dashed blue lines) of (a). Within these contour lines, the degeneracy condition is fulfilled up to a finite tolerance in the order of $C$. This procedure predicts the position of the resonances accurately, except for the fact that one out of two ``corridors'' enclosed by the contour lines contains no resonance.
}\label{fig:N2spectrum}
\end{figure*}

Figure \ref{fig:N2spectrum}(a) visualizes condition \eqref{eq:N2degencond}, by quantifying the deviation of $\mu$ from the nearest $n$-photon resonance condition, in the $F$-$\omega_0$-plane; $\mu$ is obtained by numerically solving Eq. \eqref{eq:N1FloqProbl}. As expected, the patterns in this plot reproduce the shape of the entanglement resonances. This is verified by Figure \ref{fig:N2spectrum}(b), which shows the same data as in the top left panel of Figure \ref{fig:N2_2Dplots}, superimposed by contour lines extracted from Figure \ref{fig:N2spectrum}(a). The contour values are chosen such that the lines form corridors in which the deviation from degeneracy is small (i.e., less than the interaction strength $C$). Thereby, we obtain an accurate description of the position and shape of the resonances. Note, however, that the approximate degeneracy of quasi-energies is only a necessary, but not a sufficient criterion for entanglement resonance, since one out of two corridors in Figure \ref{fig:N2spectrum}(b) contains no resonance. The absence of resonances at these degeneracies will be discussed in Sec.~\ref{ssec:N2detailedexpl}.

\begin{figure*}[tb]
	\includegraphics[width=.90\textwidth]{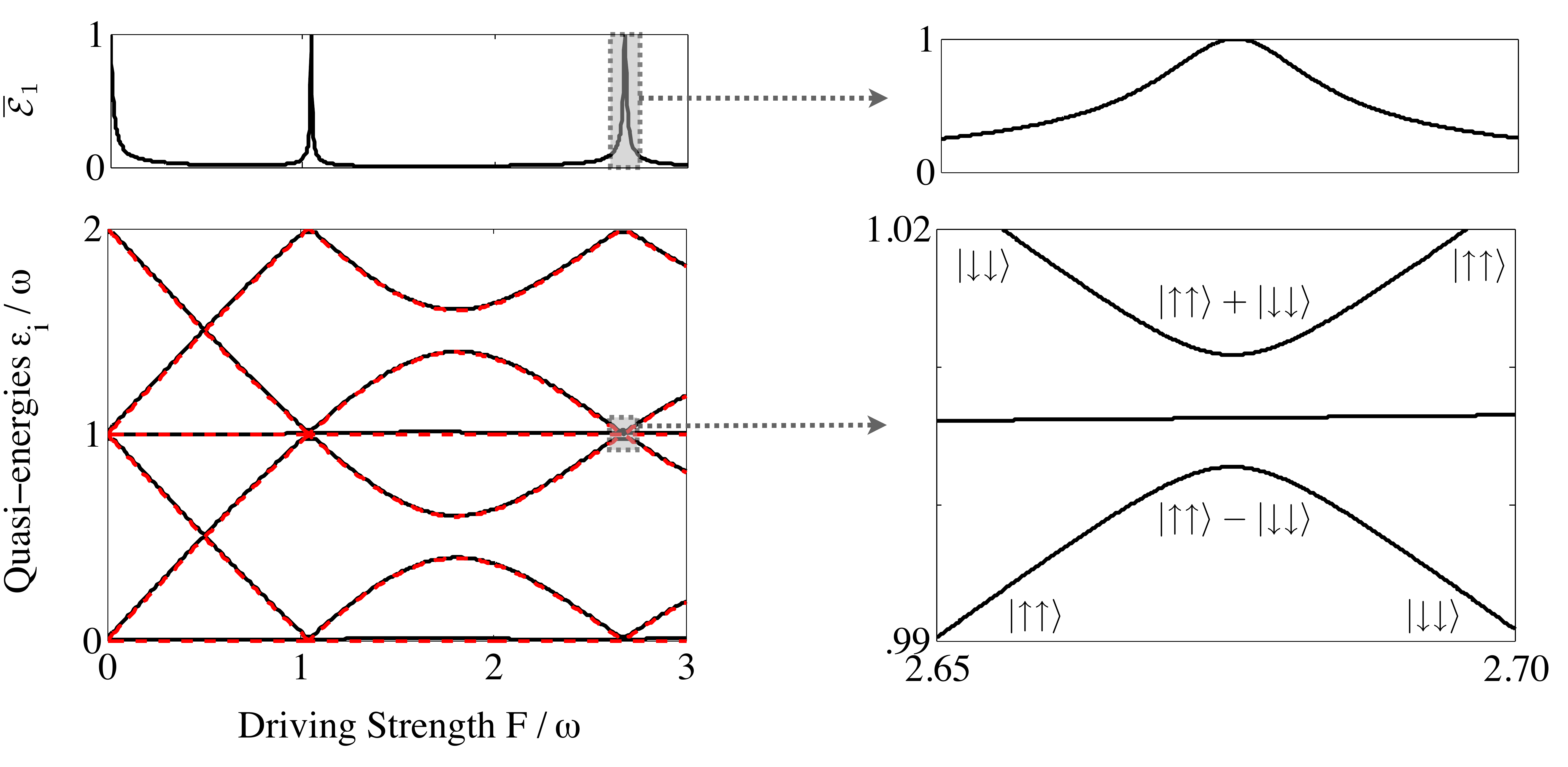}
\caption{Section through the parameter space of Figure~\ref{fig:N2spectrum}, at fixed qubit level splitting $\omega_0 = \omega$. Bottom left panel: two Floquet zones of the quasi-energy spectrum; both the levels of the weakly interacting system (black solid lines) and their non-interacting approximations (red dashed lines) are plotted. The two sets are virtually identical on this scale, due to the weak interaction $C=0.02\,\omega$. The panel above verifies that Floquet state entanglement $\overline{\mathcal{E}}_1$ peaks precisely in vicinity of (near) degeneracies of the spectrum.
Right panels: Magnification of a resonant region, revealing an avoided level crossing in the quasi-energy spectrum, on a scale in the order of $C$ (bottom, only levels of the interacting system are plotted). $\overline{\mathcal{E}}_1$ peaks precisely at its center (top).
The character of the repelling levels is indicated by a label in spin-$1/2$ notation; e.g., as explained in the text, the upper level continuously changes its character from $\ket{\Phi_2(t)}$, cf. Eq.~\eqref{eq:noninteractingFS}, which is locally unitarily equivalent to $\ket{\downarrow\downarrow}$, to $\ket{\Phi_1(t)}$, which is equivalent to $\ket{\uparrow\uparrow}$. Note that the flat level $\varepsilon_3$ is not involved in the avoided crossing scenario, since the matrix elements $|c_{13}|$ and $|c_{23}|$ describing its coupling to the other states, cf. Eq. \eqref{eq:Cmatrixel}, are vanishing here.
}\label{fig:N2section}
\end{figure*}

Figure~\ref{fig:N2section} analyzes the situation in more detail. It shows a section through the parameter plane at fixed qubit energy $\omega_0=\omega$. Both the Floquet spectra of the weakly interacting system at $C=0.02\,\omega$ (lower panels, solid black lines), and of the non-interacting system at $C=0$ (dashed red lines) are plotted, along with the entanglement $\overline{\mathcal{E}}_1$ (upper panels).
As expected, entanglement peaks in the vicinity of near-degeneracies of the spectrum. The magnification of a crossing region (right panels) reveals that the interacting levels $\varepsilon_1$ and $\varepsilon_2$ avoid to cross, with a minimal energy distance on the scale of the interaction strength $C=0.02\,\omega$. Such avoided crossings of quasi-energies are, in fact, the key to explain the resonant behavior of entanglement: Far away from their center, the Floquet states $\ket{\Phi_{1}(t)}$ and $\ket{\Phi_{2}(t)}$ of the weakly interacting system are well described by the separable ``unperturbed'' states $\ket{\phi_\pm(t)}\otimes\ket{\phi_\pm(t)}$. At the center of the avoided crossing, however, they become balanced superpositions of these product states (in complete analogy to avoided crossings in spectra of autonomous quantum systems \cite{Cohen-Tannoudji:1977}):
\begin{eqnarray*}
	\ket{\Phi_{1}(t)} = \frac{1}{\sqrt{2}} (\ket{\phi_+(t)}\otimes\ket{\phi_+(t)}+ \ket{\phi_-(t)}\otimes \ket{\phi_-(t)}), \\
	\ket{\Phi_{2}(t)} = \frac{1}{\sqrt{2}} (\ket{\phi_+(t)}\otimes\ket{\phi_+(t)} - \ket{\phi_-(t)}\otimes \ket{\phi_-(t)}).
\end{eqnarray*}
These are maximally entangled states at all times $t$, as can be seen, once again, by application of the local unitary transformation $U_\phi(t) \otimes U_\phi(t)$, Eq. \eqref{eq:lu}.

Precisely the same mechanism occurs at higher order resonances, i.e., when condition \eqref{eq:N2degencond} is fulfilled for $n>0$. Then, $\ket{\Phi_1(t)}$ is degenerate with $e^{i n \omega t}\ket{\Phi_2(t)}$. Hence, at the center of an avoided crossing between these levels, we have
\begin{equation*}
	\frac{1}{\sqrt{2}} (\ket{\phi_+(t)}\otimes\ket{\phi_+(t)} \pm e^{i n \omega t} \ket{\phi_-(t)}\otimes \ket{\phi_-(t)})
\end{equation*}
as Floquet states of the interacting system, which again are maximally entangled states at all times $t$ \footnote{To see this, one has to modify $U_\phi(t)$ of Eq. \eqref{eq:lu} to $U_\phi(t) = \ket{\uparrow}\bra{\phi_+(t)} + e^{-i n \omega t/2}\ket{\downarrow}\bra{\phi_-(t)}$.}.

From our perturbative analysis, one can also understand what happens in the case of larger qubit-qubit coupling strength: As $C$ grows, the avoided crossings widen and migrate from their original position, resulting in broader and slightly shifted entanglement resonances that eventually overlap, and form the rich patterns observed in the middle and bottom row of Figure~\ref{fig:N2_2Dplots}.

To summarize our discussion so far, we have identified the central mechanism the underlies entanglement resonances: Whenever the weak interaction between the qubits lifts a degeneracy of the non-interacting two-qubit Floquet spectrum, it locally strongly couples the anti-crossing Floquet states and transforms them from separable into maximally entangled states. For clarity, let us emphasize that this mechanism is not unique to avoided crossings in Floquet spectra, but similarly occurs for eigenstates of weakly interacting, \textit{autonomous} quantum systems, whenever the corresponding energy levels cross under variation of a static control parameter. This phenomenon has been discussed, e.g., for spins chains \cite{Karthik:2007,Bruss:2005}. 
However, control of entanglement (and more generally, of $N$-body interaction) by oscillating fields, as suggested here, is much more versatile, since the rest class structure of the Floquet spectrum leads to \textit{multi-photon} resonance conditions (alike \eqref{eq:N2degencond}), and thereby allows to address the quantum many-particle system through a multitude of side-bands, which are absent in static control scenarios, and which might experimentally be much easier to access.


\subsection{Detailed analysis of entanglement resonances}\label{ssec:N2detailedexpl}

Based on our above, qualitative explanation of the entanglement resonances, we now develop a better understanding of their position and line shape in parameter space. This involves two steps:
\begin{enumerate}
\item First, we derive (approximate) expressions for the single qubit quasi-energy $\mu$ as a function of the driving amplitude $F$ and of the qubit energy splitting $\omega_0$. This allows to parametrize the degeneracy condition \eqref{eq:N2degencond}, and thus provides an analytical description of the contour lines of Figure~\ref{fig:N2spectrum}(b).
\item Next, we study the coupling strength between Floquet levels, i.e., the width of avoided crossings which generate the entanglement resonances, and investigate why it vanishes within those corridors that contain no resonance.
\end{enumerate}

The first step can be achieved only within certain approximations, since the single driven qubit problem, Eq.~\eqref{eq:N1FloqProbl}, has no closed solution for the monochromatic driving scheme of Eq.~\eqref{eq:N1Hamiltonian}. In the regime of weak driving, $F \ll \omega_0 $, and small detuning, $\omega_0 \approx \omega$, the rotating wave approximation (RWA) can be applied \cite{Cohen-Tannoudji:1998}. Note that these conditions are fulfilled only in a small fraction of the parameter plane shown in Figs. \ref{fig:N2_2Dplots} and \ref{fig:N2spectrum}, namely in the vicinity of $(F=0,\omega_0=\omega)\equiv P_0$. The RWA consists of neglecting the ``counter-rotating'' terms of the driving field, i.e., in replacing
\begin{equation*}
	F\cos(\omega t) \sigma_x = \frac{F}{2}\left( e^{i\omega t} \sigma_+ + e^{-i\omega t} \sigma_+ + e^{i\omega t} \sigma_- + e^{-i\omega t} \sigma_- \right)
\end{equation*}
by
\begin{equation*}
	\frac{F}{2}\left( e^{-i\omega t} \sigma_+ + e^{i\omega t} \sigma_-\right).
\end{equation*}
This way, the Floquet Hamiltonian \eqref{eq:Floq_Hamiltonian2} decomposes into decoupled $2\times 2$ blocks (each spread out over two Fourier components) and can be diagonalized by hand
, leading to quasi-energies
\begin{equation}\label{eq:mu_RWA}
	\mu_\pm^\textrm{RWA} = \pm \frac{1}{2} \left( \omega + \sqrt{(\omega-\omega_0)^2 + F^2}\right).
\end{equation}
With this, the degeneracy condition \eqref{eq:N2degencond} turns into
\begin{equation}\label{eq:N2degencondRWA}
	\sqrt{\left(\omega-\omega_0\right)^2 + F^2} = n\omega /2  \qquad (n\in \mathbb{N}_0).
\end{equation}
This defines a circles of radius $n \omega/2$ around $P_0$ in the $F$-$\omega_0$-plane. Hence, for $n=0$, condition \eqref{eq:N2degencondRWA} predicts a point-like resonance at $P_0$. It is hardly visible in Figure~\ref{fig:N2spectrum}(b), but appears broader in the middle row of Figure~\ref{fig:N2_2Dplots}. For $n=1$, the condition describes a circular corridor of radius $\omega/2$ around $P_0$, as visible in Figure~\ref{fig:N2spectrum}(b). Beyond this corridor, the RWA is no longer justified, and, accordingly, the location of all other corridors in Figure~\ref{fig:N2spectrum}(b) deviates quite strongly from the circular shape. Only in the weak driving limit of higher order resonances, i.e., at $F\ll \omega$ and $\omega_0\approx 2 \omega$, $\omega_0\approx 3 \omega$, etc., does RWA provide a reasonably accurate description. In these parameter regions, multi-photon transitions are resonant with the unperturbed qubit transition frequencies, and the Floquet Hamiltonian again effectively decouples into $2\times 2$ blocks, with the same degeneracy condition \eqref{eq:N2degencondRWA} \cite{Shirley:1965}.

\begin{figure}[tb]
\includegraphics[width=0.45\textwidth]{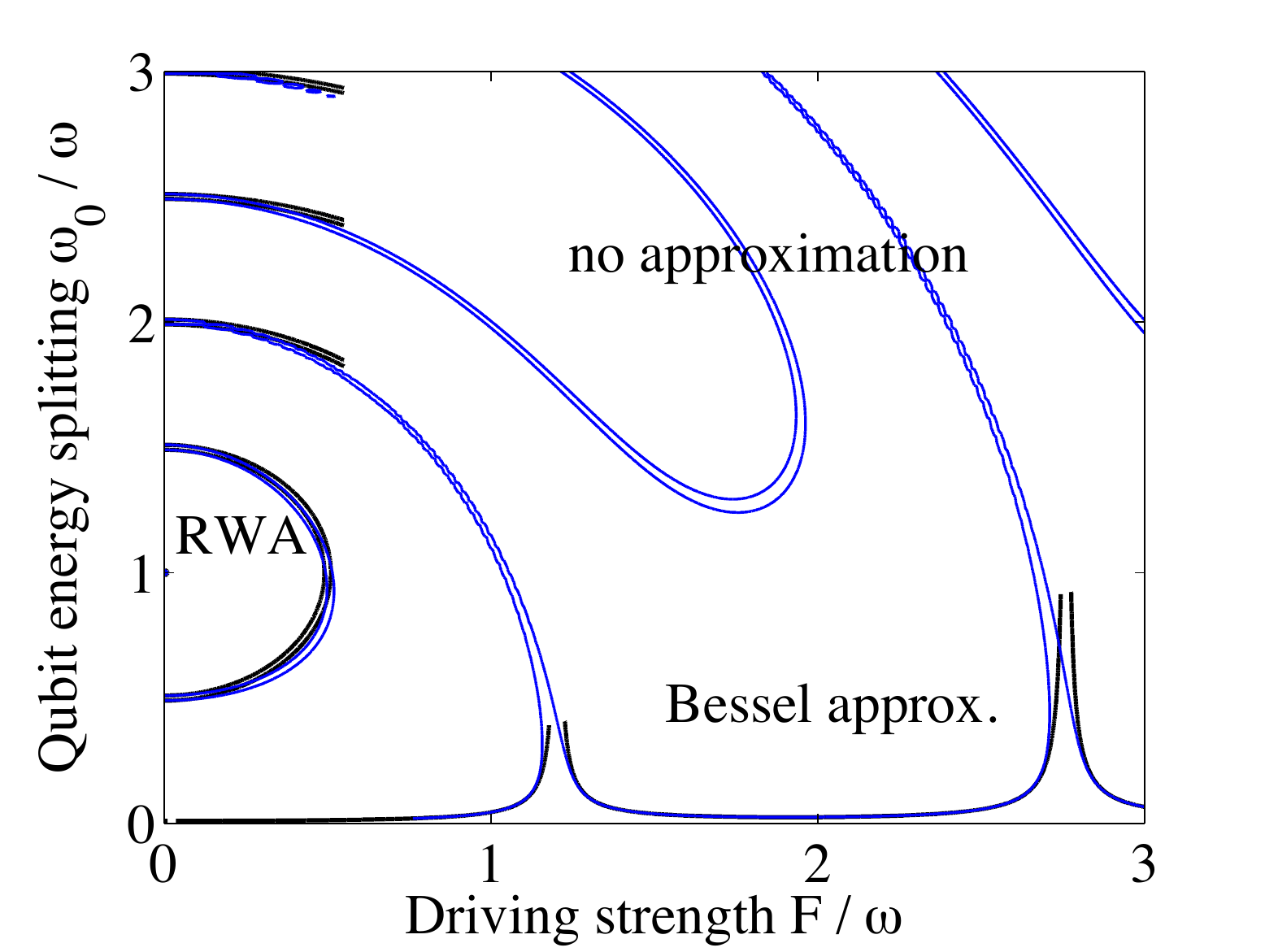}
\caption{Illustration of two analytical approximations to the contour lines of Figure~\ref{fig:N2spectrum}(b). For weak driving amplitude $F$, a description in terms of the RWA holds, explaining the circular shape of contour lines in this area, cf. Eq.~\eqref{eq:N2degencondRWA}. For small qubit energies $\omega_0 \ll F$, the single qubit quasi-energy $\mu$ can be approximated by a Bessel function, cf. Eq.~\eqref{eq:mu_smallomega0}. The two approximations (bold black lines) render correctly the shape of the resonances in the respective regions of parameter space. For large qubit energy \textit{and} strong driving, no analytical description is available, and one has to resort to the numerical determination of $\mu$ (thin blue lines).}\label{fig:contourRWA}
\end{figure}

For large driving strength $F$, the Floquet Hamiltonian strongly couples more than two frequency components, and can no longer be approximated by decoupled $2\times 2$ blocks. This leads to a failure of the RWA resonance condition \eqref{eq:N2degencondRWA}. The Floquet states acquire a large number of non-vanishing Fourier components, and this makes it difficult to analytically explain the line shapes of the resonances. However, at least in the regime of small qubit energies $\omega_0 \ll \omega$, an approximation can be found that is valid for arbitrary driving strengths \cite{Shirley:1965}: It is given in terms of the zeroth order Bessel function of the first kind, $J_0$:
\begin{equation}\label{eq:mu_smallomega0}
	\mu^{\omega_0\ll\omega} = \frac{\omega_0}{2} J_0\left(2 \frac{F}{\omega}\right).
\end{equation}
This expression vanishes at $\omega_{0}=0$, consistent with the degeneracy observed in Figure~\ref{fig:N2spectrum} in the limit $\omega_0 \rightarrow 0$ . Furthermore, it accurately describes the resonance positions close to the $F$-axis, which coincide with the zeros of $J_0$.
In Figure~\ref{fig:contourRWA}, the exact contour lines are plotted along with the approximate expressions \eqref{eq:N2degencondRWA} and \eqref{eq:mu_smallomega0}, to illustrate their respective regions of validity.

The widths of entanglement resonances are given by the widths of the associated avoided crossings. According to Eq.~\eqref{eq:Cmatrixel_theory}, this width is determined, at first order perturbation theory, by the matrix elements $c_{ij}$ of the participating Floquet states $\ket{\Phi_{i}(t)}$ and $\ket{\Phi_{j}(t)}$ with respect to the qubit-qubit interaction $H_\textrm{qq}$:
\begin{equation}\label{eq:Cmatrixel}
	c_{ij} = \frac{\omega}{2\pi} \int_0^{2\pi/\omega} dt \, \braket{\Phi_i(t) | C (\sigma_{+}^{(1)} \sigma_{-}^{(2)} + \sigma_{-}^{(1)} \sigma_{+}^{(2)}) | \Phi_j(t)}.
\end{equation}
Thus, at first order, the width of the resonances scales linearly in the qubit-qubit interaction strength $C$ and is bounded by $2|c_{ij}|\le 2C$. This is why the contour lines in Figure~\ref{fig:N2spectrum}(b) are chosen such that they enclose regions in which the unperturbed levels come closer than $2C$, thereby defining an upper limit for the width of the resonances.

\begin{figure}[tb]
\includegraphics[width=0.55\textwidth]{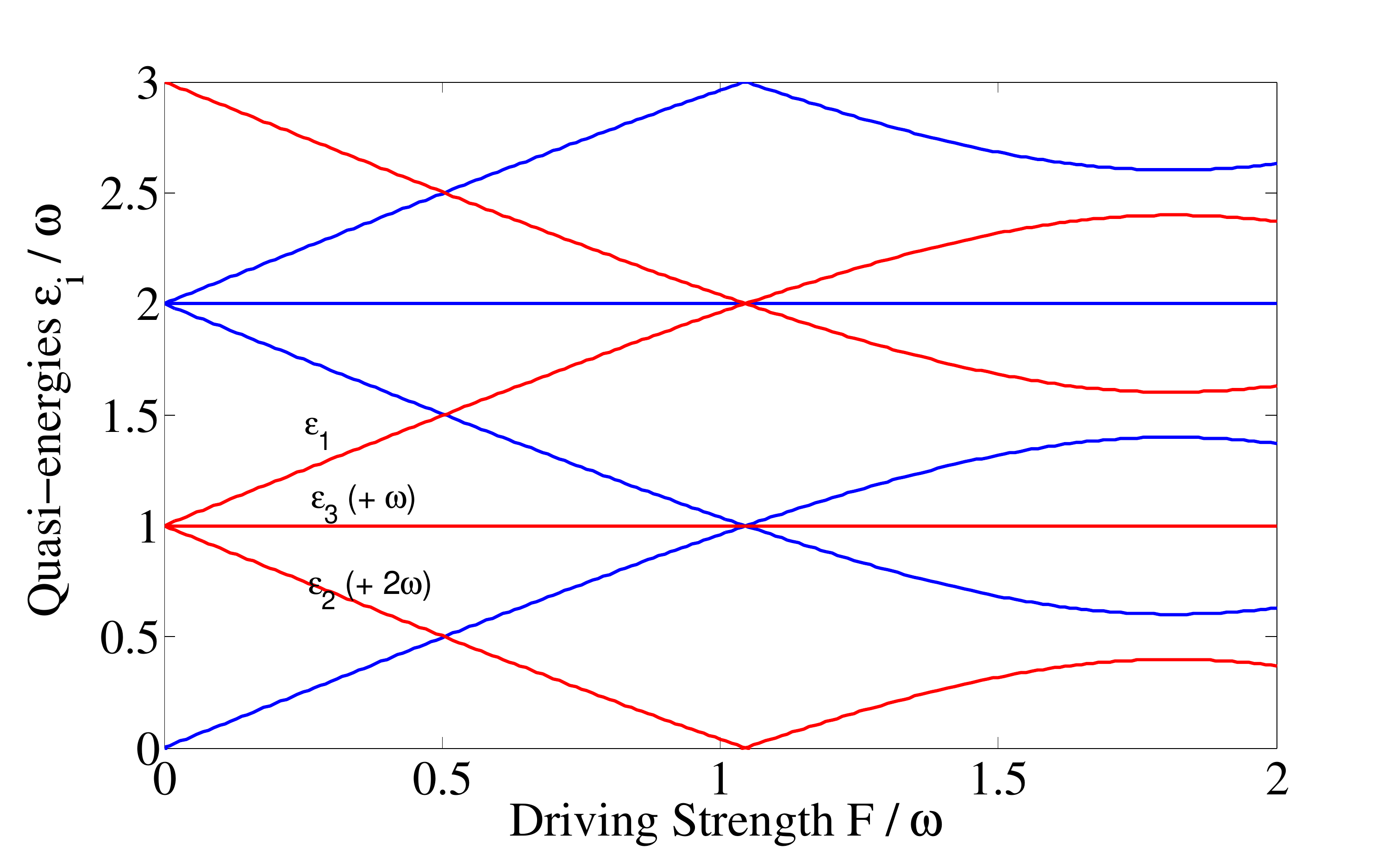}
\caption{Three Floquet zones of the Floquet spectrum of two monochromatically driven qubits without qubit-qubit interaction ($C=0$), at $\omega_0=\omega$. Levels are colored red and blue, according to their symmetry with respect to the generalized parity $\mathcal{S}_{P2}$, Eq. \eqref{eq:SP2}. Since the symmetry is retained in the presence of qubit-qubit interaction, level crossings between different classes do not turn into avoided crossings in presence of a finite interaction ($C>0$). This explains the absence of entanglement resonances in one out of two corridors in Figure~\ref{fig:N2spectrum}(b). For clarity, we emphasize that the level distortion around $F=1.5...2\,\omega$ is \textit{not} related to the interaction-induced avoided crossings, which lead to entanglement resonances (after all, the plot is obtained for $C=0$). This distortion is rather a property of the single driven qubit spectrum for strong driving, indicating the departure from the RWA regime.}\label{fig:N2symmplot}
\end{figure}

The absence of entanglement resonances in some of the such defined corridors, as observed in Figure\ref{fig:N2spectrum}(b), can be traced back to a generalized parity of the single qubit Floquet problem \eqref{eq:N1FloqProbl} \cite{Grossmann:1991,Buchleitner:2002,Breuer:1988,Braak:2011}. The operator $\mathcal{S}_P$ associated with this parity symmetry acts on the Floquet Hilbert space of a single driven qubit and reads
\begin{equation}\label{eq:SP}
	\mathcal{S}_P \ket{\phi(t)} = \sigma_z \ket{\phi(t+\pi/\omega)}.
\end{equation}
Since it fulfills $(\mathcal{S}_{P})^{2}=1$ and commutes with the Floquet Hamiltonian $h(t)-i\partial_t$, the single qubit Floquet states $\ket{\phi_+ (t)}$ and $\ket{\phi_- (t)}$ are eigenstates of $\mathcal{S}_P$, with eigenvalues $\pm1$, respectively. In the two qubit Floquet Hilbert space, the corresponding parity $\mathcal{S}_{P2}$ reads
\begin{equation}\label{eq:SP2}
	\mathcal{S}_{P2} \ket{\Phi(t)} = \sigma_z^{1} \sigma_z^{2} \ket{\Phi(t+\pi/\omega)}.
\end{equation}
The non-interacting Floquet state $\ket{\Phi_1(t)}=\ket{\phi_+(t)}^{\otimes 2}$, cf. Eq. \eqref{eq:noninteractingFS}, has parity $+1$ with respect to $\mathcal{S}_{P2}$, because
\begin{equation}
	\mathcal{S}_{P2} \ket{\phi_+(t)}^{\otimes 2} = (\mathcal{S}_{P} \ket{\phi_+(t)})^{\otimes 2} = (+1)^2 \ket{\phi_+(t)}^{\otimes 2}.
\end{equation}
Likewise, the second non-interacting Floquet state $\ket{\Phi_2 (t)}=\ket{\phi_-(t)}^{\otimes 2}$ has positive parity, whereas $\ket{\Phi_3(t)}=\frac{1}{\sqrt{2}} ( \ket{\phi_+(t)} \otimes \ket{\phi_-(t)} + \ket{\phi_-(t)} \otimes \ket{\phi_+(t)} )$ has negative parity. Since also the qubit-qubit interaction term that we considered so far (the $\sigma_+\sigma_-$ ``excitation exchange interaction'') commutes with $\mathcal{S}_{P2}$, levels of different parity are not coupled by this interaction. Consequently, driving-induced degeneracies of two-qubit Floquet states of opposite parity are \textit{not} lifted by a non-vanishing qubit-qubit interaction strength $C$. Only levels of equal parity are coupled by $H_{qq}$ and can thus give rise to an entanglement resonance.

To illustrate the above discussion, three Floquet zones of the non-interacting quasi-energy spectrum are shown in Figure \ref{fig:N2symmplot}, with levels colored according to their parity. As explained in Sec.~\ref{ssec:Floqtheory}, the rest class structure of Floquet states results in an $\omega$-periodicity of the Floquet spectrum. This implies that for each Floquet state $\ket{\Phi_i(t)}$ of quasi-energy $\varepsilon_i$, there is a homologue Floquet state $\ket{\Phi_i(t)} e^{i \omega t}$ of quasi-energy $\varepsilon_i-\omega$ in the neighboring Floquet zone. If $\ket{\Phi_i(t)}$ is symmetric with respect to $\mathcal{S}_{P2}$, then $\ket{\Phi_i(t)}e^{i\omega t}$ is antisymmetric, and vice versa, since of $e^{i \omega (t+\pi/\omega)} = - e^{i \omega t } $; i.e., the parity of a level switches from Floquet zone to Floquet zone. This implies that an avoided crossing of levels is symmetry-forbidden whenever the degeneracy condition $\eqref{eq:N2degencond}$ is fulfilled for odd $n$, and explains why precisely every other corridor in Figure~\ref{fig:N2spectrum}(b) contains no entanglement resonance.

\subsection{Bi-chromatic, saw-tooth, and $\delta$-kicked driving}\label{ssec:N2bichromsawtooth}
\begin{figure*}[tb]
\begin{tabular}{l l}
	(a) $F'=F/2$ & (b) $F'=F$\\
	\includegraphics[width=.45\textwidth]{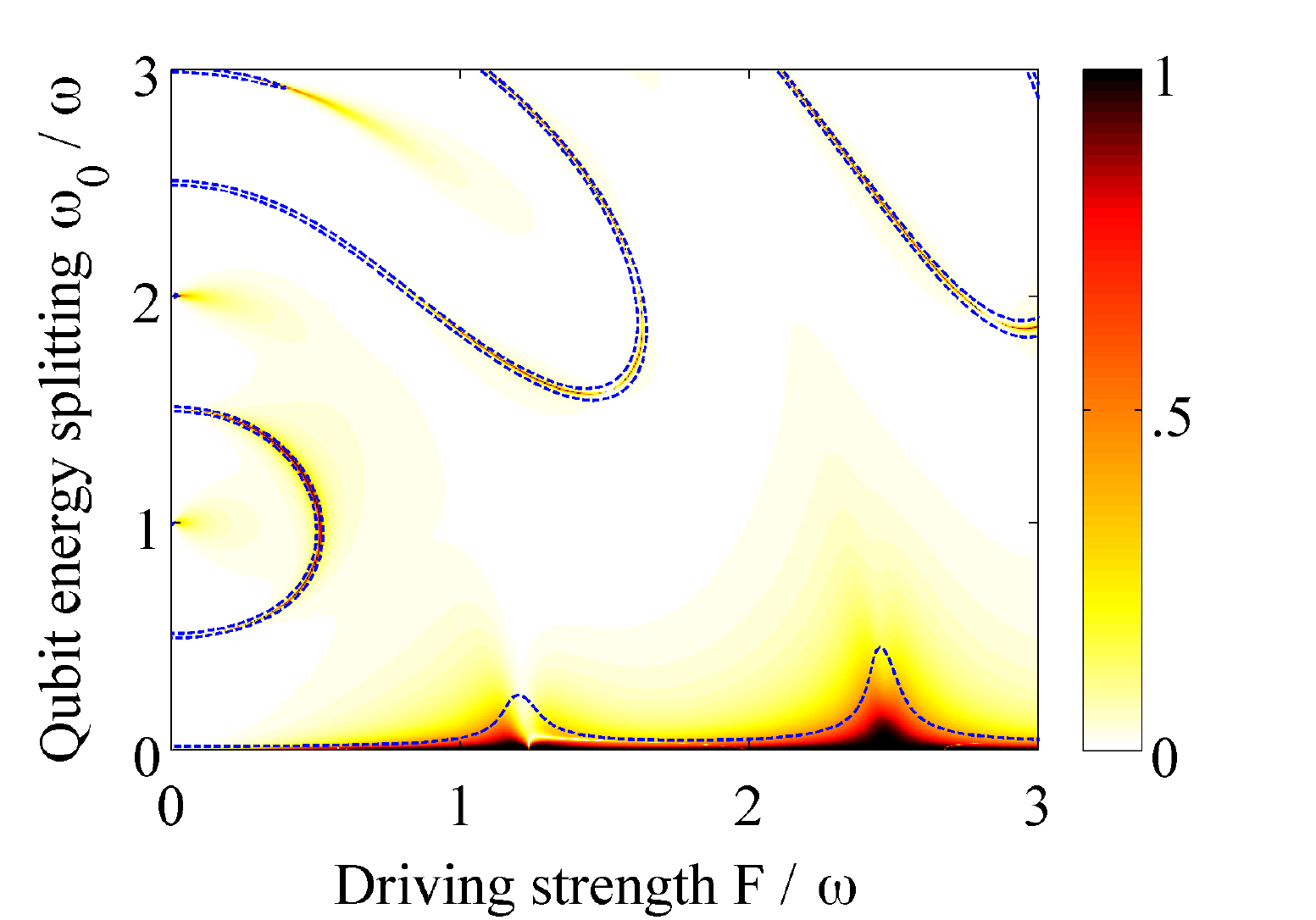} & \includegraphics[width=.45\textwidth]{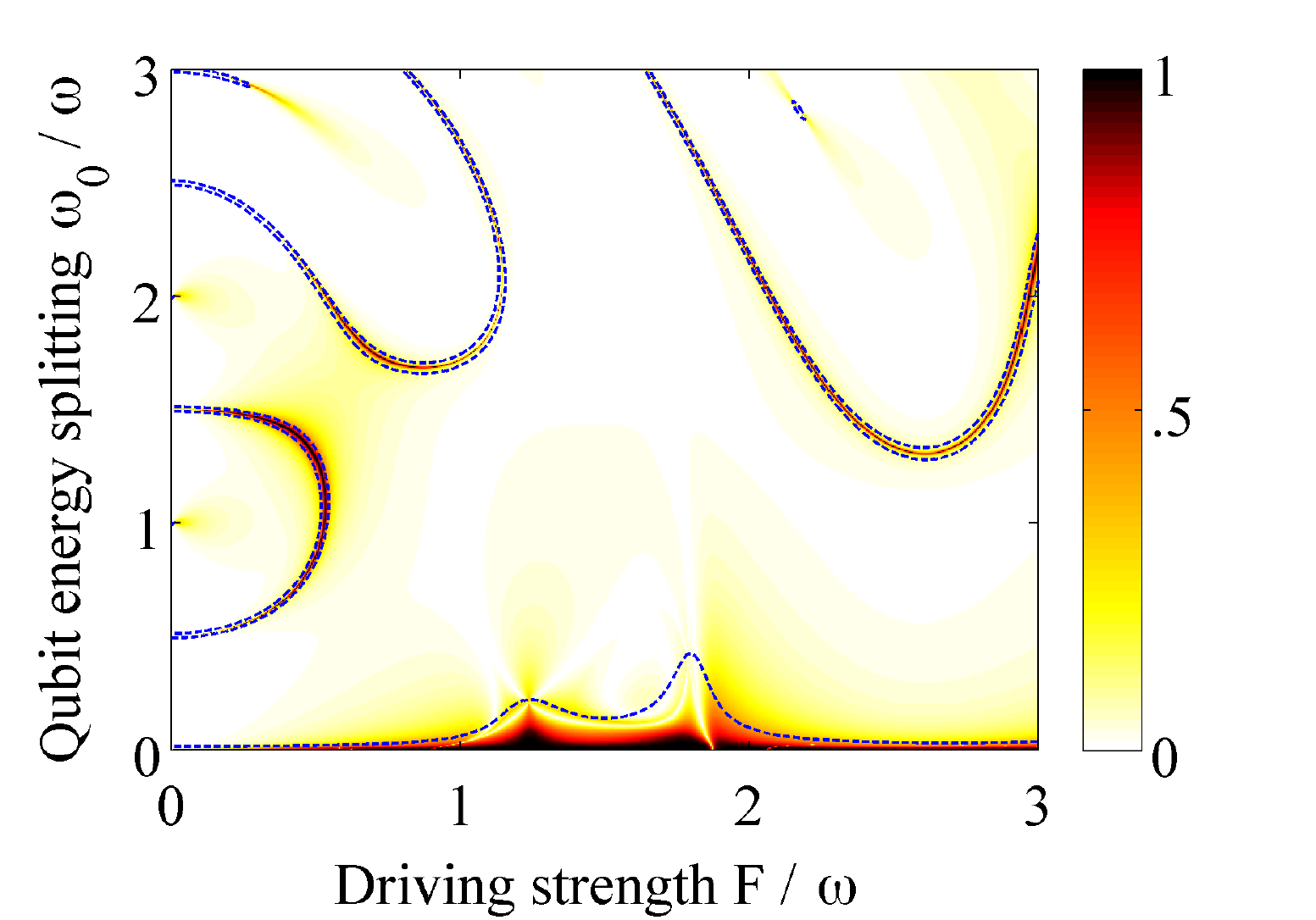}
\end{tabular}
\caption{Entanglement $\overline{\mathcal{E}}_1$ of the least entangled symmetric Floquet state of two weakly coupled qubits ($C=0.02\,\omega$) under bi-chromatic driving, i.e., $f(t)=F \cos(\omega) t + F' \cos(2\omega t)$. (a) $F'=F/2$. (b) $F'=F$. The predictions (dashed blue lines) are contour lines derived from the degeneracy condition \eqref{eq:N2degencond}, as explained in the caption of Figure~\ref{fig:N2spectrum}(b).}
\label{fig:N2bichrom}
\end{figure*}

To underpin our discussion of symmetry-suppressed resonances, we consider a bi-chromatic driving scheme, $f(t)=F \cos(\omega) t + F' \cos(2\omega t)$, which is easily generated in laboratories by using the second harmonic field mode \cite{Shen:1984}. This scheme breaks the generalized symmetry $\mathcal{S}_{P}$. Therefore, all levels couple to each other via the qubit-qubit interaction, and we expect an entanglement resonance \textit{whenever} the degeneracy condition \eqref{eq:N2degencond} is fulfilled. Figure~\ref{fig:N2bichrom} shows the entanglement $\mathcal{E}_{1}$ of the first Floquet state, and confirms that now \textit{all} corridors defined by the degeneracy condition enclose an entanglement resonance. (We omit plotting $\mathcal{E}_{2}$ and $\mathcal{E}_{3}$ for the bi-chromatic case, since, as in the monochromatic case, $\mathcal{E}_{2}$ is very similar to $\mathcal{E}_{1}$, while $\mathcal{E}_{3}$ is maximal almost everywhere.) Compared to the monochromatic case, the corridors are parametrized differently here, since they derive from the single qubit quasi-energy $\mu$ under bi-chromatic driving, which is different from the monochromatic case. Still, our prediction scheme based on the degeneracy condition accurately describes the resonances.

\begin{figure}[tb]
\begin{tabular}{l l}
	(a) Sawtooth driving & (b) Driving by periodic $\delta$-kicks\\
 \includegraphics[width=.45\textwidth]{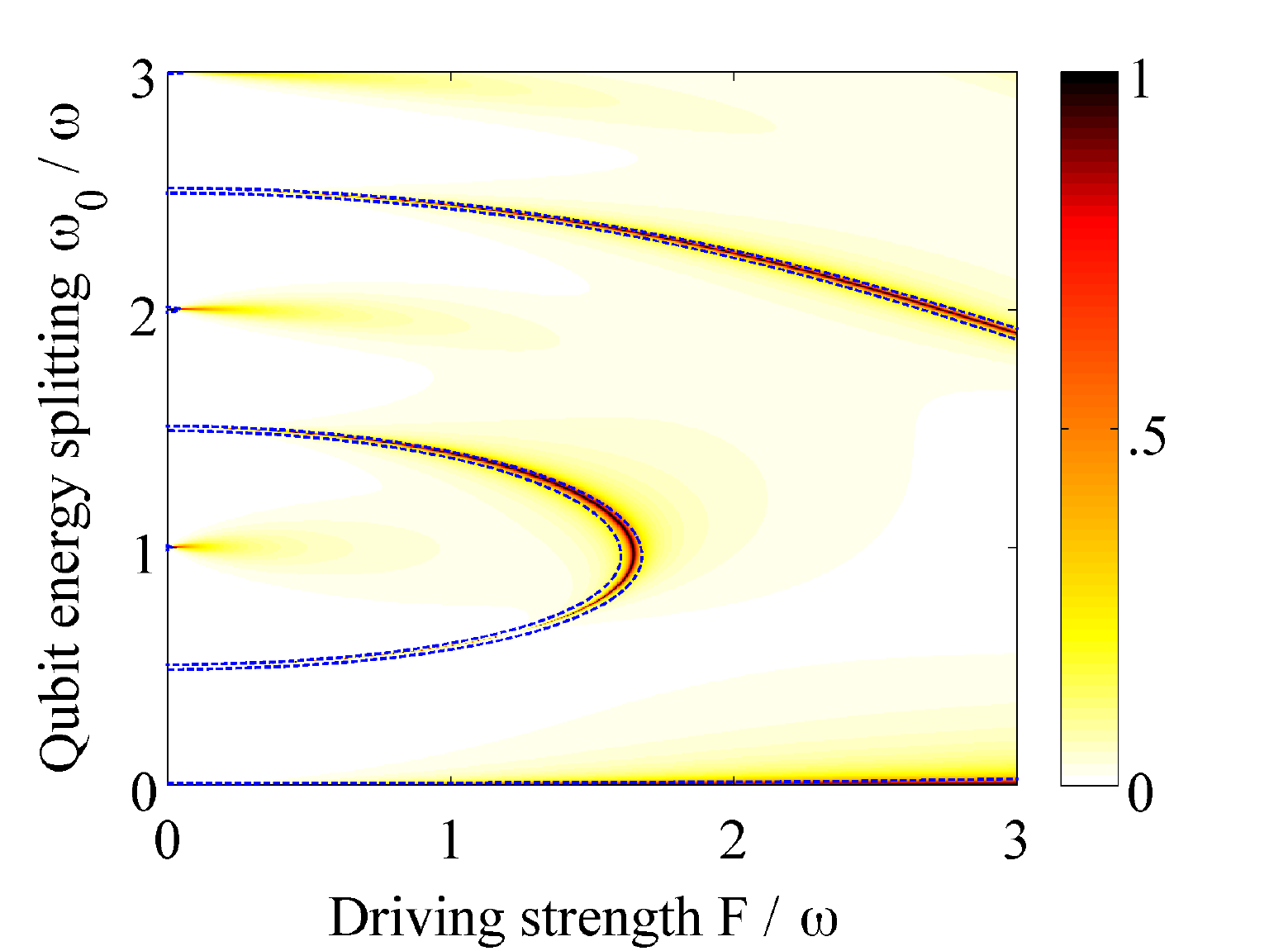} & \includegraphics[width=.45\textwidth]{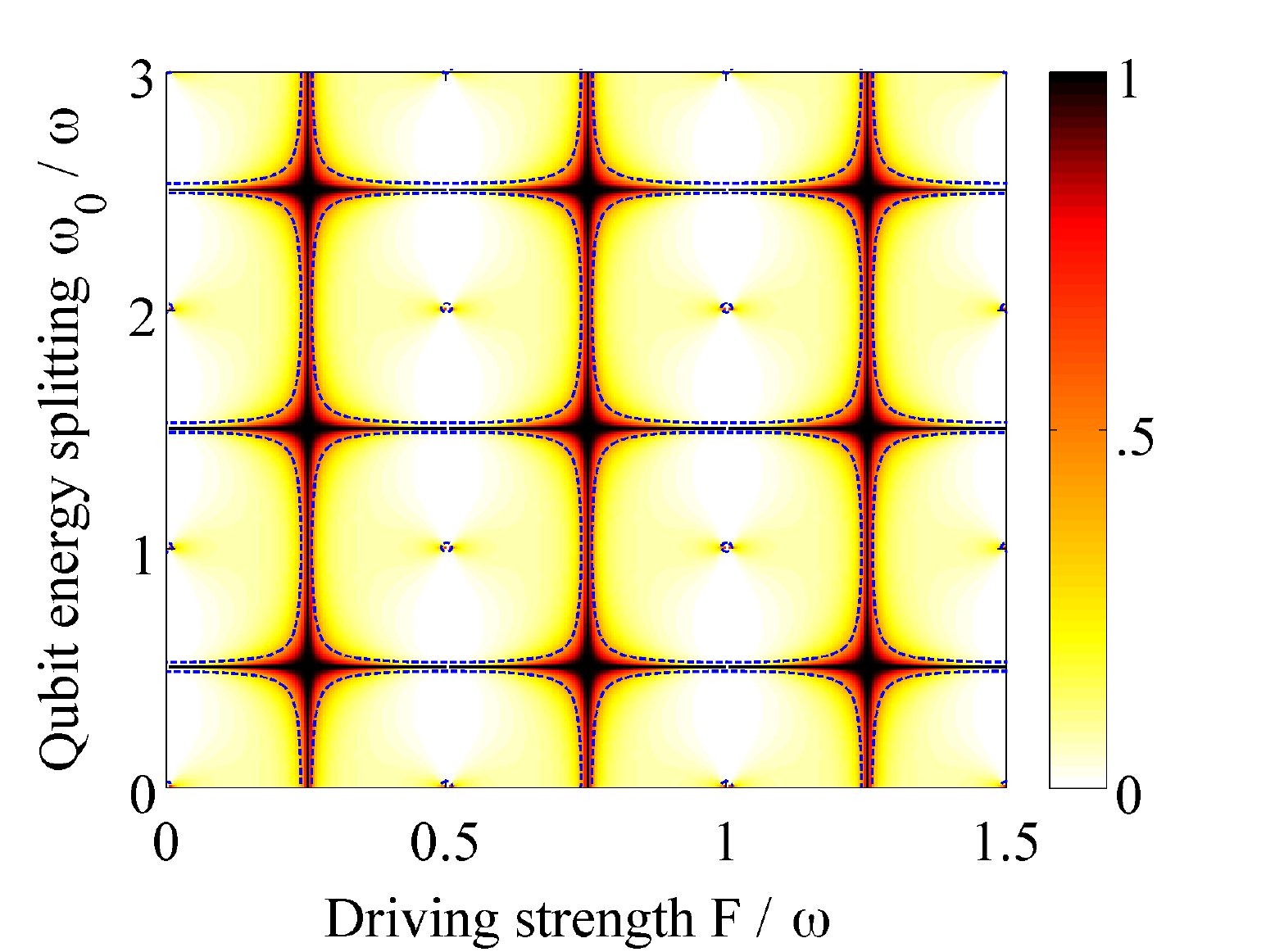}
 \end{tabular}
\caption{Entanglement $\overline{\mathcal{E}}_1$ of two weakly coupled qubits, driven by (a) a saw-tooth profile $f(t)=F \cdot \left[\frac{t}{T} \textrm{ mod } 1 - \frac{1}{2}\right]$, (b) periodic $\delta$-kicks, $f(t)=FT \cdot \delta(t\mod T)$. The predictions (dashed blue lines) are obtained from the exact analytical expressions \eqref{eq:SawtoothExplicitQE} and \eqref{eq:deltakickExplicitQE}, respectively.}
\label{fig:N2sawtooth}
\end{figure}

Next, as an example of analytically solvable single qubit dynamics, we consider a saw-tooth driving profile $f(t)=F \cdot \left[\frac{t}{T} \textrm{ mod } 1 - \frac{1}{2}\right]$ that periodically ramps up the driving amplitude from $-F/2$ to $+F/2$, with period $T=2\pi/\omega$. Thus, the single qubit Hamiltonian is
\begin{equation}\label{eq:N1HamiltonianSAWTOOTH}
	h(t) = \frac{\omega_0}{2} \sigma_z + F \cdot \left[\frac{t}{T} \textrm{ mod } 1 - \frac{1}{2}\right] \sigma_x.
\end{equation}
Note that this describes a periodic repetition of a Landau-Zener scenario (in which the diabatic states are the eigenstates of $\sigma_x$, and $\omega_0$ plays the role of a coupling strength between the diabatic states). Based the analytical solution of a single Landau-Zener transition \cite{Akulin:2006}, we derive the explicit expression 
\begin{equation}\label{eq:SawtoothExplicitQE}
	\mu = \frac{1}{T} \arccos\left[ 2 \left|\, _1 F_1\left(\frac{-i \omega_0^2 T}{16 F},\frac{1}{2}, -\frac{i F T}{4} \right) \right|^2 - 1 \right]
\end{equation}
for the single qubit quasi-energy $\mu$ in \ref{sec:AppendixSawtooth}, with $_{1}F_{1}$ denoting Kummer's function \cite{Abramowitz:1964}. Inserting this into the degeneracy condition \eqref{eq:N2degencond}, one accurately predicts the positions of entanglement resonances, as illustrated in Figure~\ref{fig:N2sawtooth}(a).

In the same Appendix, we also derive the analytical expression for the quasi-energies of a qubit exposed to a train of $\delta$-kicks \cite{Hillermeier:1992}, i.e., for the driving profile $f(t)=FT \cdot \delta(t\mod T)$:
\begin{equation}\label{eq:deltakickExplicitQE}
	\mu = \frac{1}{T} \arccos\left[ \cos\left(\frac{\omega_0 T}{2}\right) \cdot \cos (FT) \right].
\end{equation}
Also here, entanglement resonances appear in the parameter regions predicted by the degeneracy condition, cf. Figure~\ref{fig:N2sawtooth}(b).

Generally speaking, one can create entanglement resonances with virtually any driving profile $f(t)$, if only the single qubit quasi-energy $\mu$ can be tuned to zero by variation of the driving parameters. 

\subsection{Variation of the interaction term}\label{sec:interactionvariation}
So far, we studied Hamiltonian \eqref{eq:Hamiltonian2} for different driving profiles $f(t)$, but fixed the interaction mechanism $H_\textrm{qq}$ between the qubits.
Based on our understanding of entanglement resonances, we expect the details of this interaction not to be decisive for the phenomenon: the corridors that define parameter regions of driving-induced near-degeneracies of the uncoupled Floquet states -- i.e., the regions where entanglement resonances may arise -- are independent of $H_\textrm{qq}$, since they derive from the degeneracy condition \eqref{eq:N2degencond}, which relies on the single qubit quasi-energy $\mu$ alone. The role of $H_\textrm{qq}$ is to lift the degeneracies of the non-interacting spectrum, and any generic two-qubit operator will achieve this. Of course, the details of the interaction determine the exact value of the coupling matrix elements $c_{ij}$ in Eq.~\eqref{eq:Cmatrixel}, and, hence, define the width of the resonances; furthermore, we have seen above that, if the interaction preserves a certain symmetry of the single qubit Floquet Hamiltonian, some degeneracies may not be lifted, i.e., some entanglement resonances may be symmetry-forbidden.

\begin{figure*}[tb]
\includegraphics[width=0.32\textwidth]{N2Phi1C002}
\includegraphics[width=0.32\textwidth]{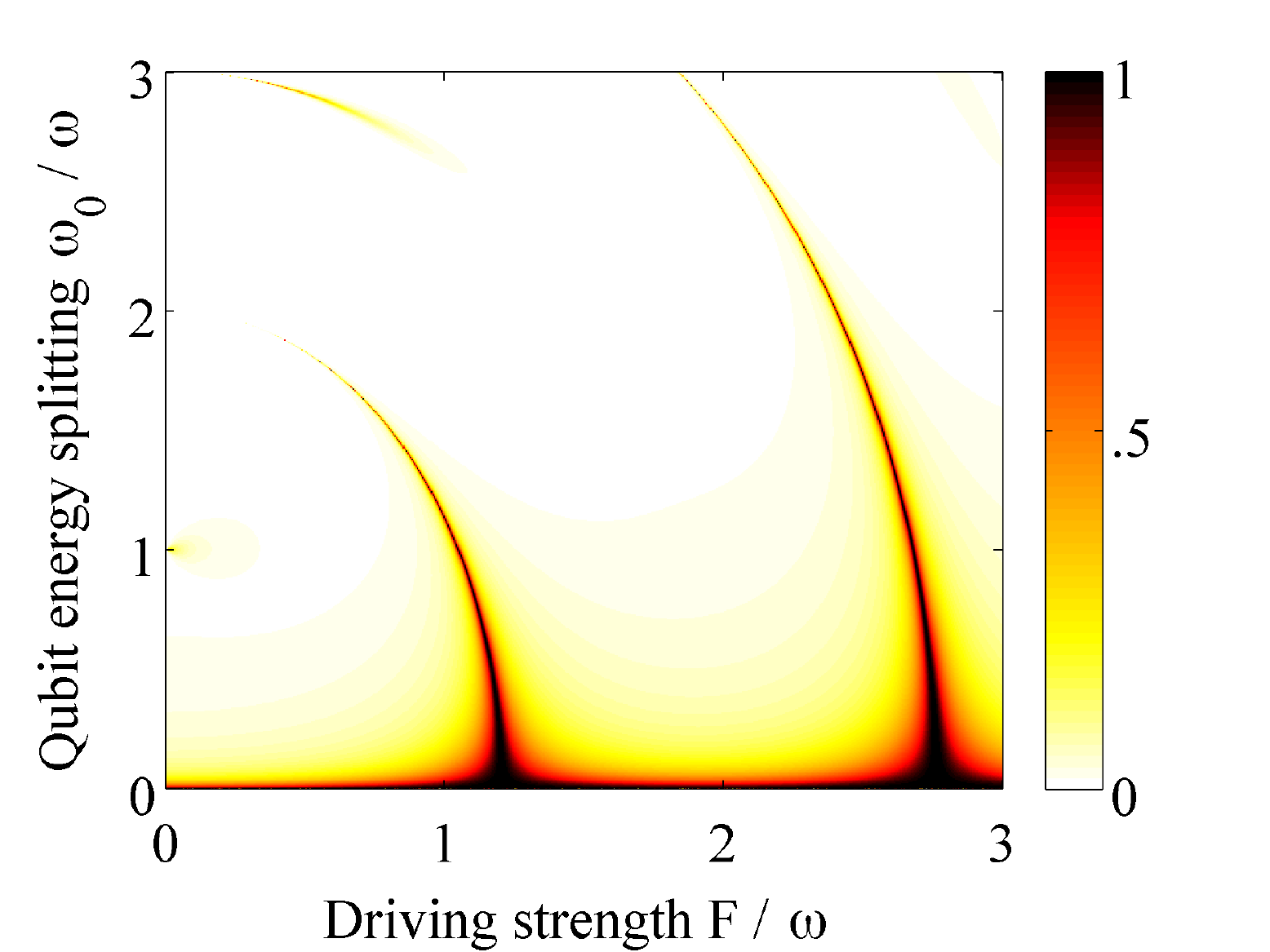}
\includegraphics[width=0.32\textwidth]{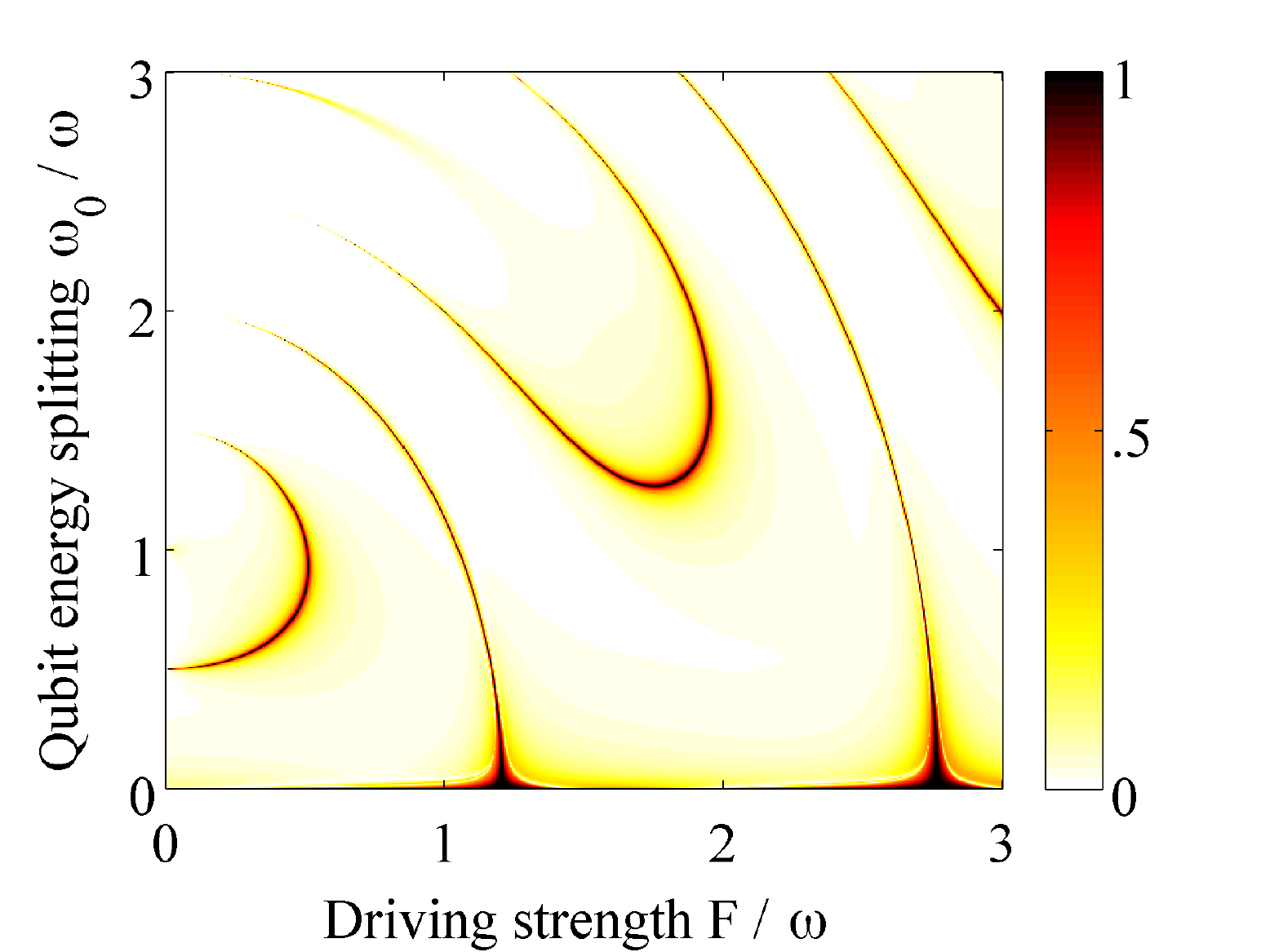}
\caption{Influence of the qubit-qubit interaction $H_\textrm{qq}$ on the Floquet state entanglement $\overline{\mathcal{E}}_1$ of two weakly coupled, monochromatically driven qubits. The left panel shows $H_\textrm{qq}=C(\sigma_{+}^{1}\sigma_{-}^{2} + \sigma_{-}^{1}\sigma_{+}^{2})$, with $C=0.02\,\omega$, as considered in the previous figures (i.e., it is identical to Figure~\ref{fig:N2spectrum}(b)). In the middle and right panels, the interaction term is replaced by $C\,\sigma_{x}^{(1)}\sigma_{x}^{(2)}$ and $C(\sigma_{x}^{(1)}+\sigma_{z}^{(1)})(\sigma_{x}^{(2)}+\sigma_{z}^{(2)})/2$, respectively. The latter interaction breaks the generalized symmetry $\mathcal{S}_{P2}$, and therefore leads to twice as many resonances.}
\label{fig:variationinteraction}
\end{figure*}
This reasoning is illustrated in Figure~\ref{fig:variationinteraction}. It shows Floquet state entanglement under monochromatic driving for different interaction mechanisms. The left plot corresponds to the excitation exchange interaction of Eq.~\eqref{eq:Hqq2}. It is replaced by $\sigma_{x}^{(1)}\sigma_{x}^{(2)}$ in the central, and by $(\sigma_{x}^{(1)}+\sigma_{z}^{(1)})(\sigma_{x}^{(2)}+\sigma_{z}^{(2)})/2$ in the right plot \footnote{If the Pauli matrices refer to a physical spin-1/2 particle, and not to an abstract two-level system, the different interactions have the following physical meaning: $\sigma_{x}^{(1)}\sigma_{x}^{(2)}$ and $(\sigma_{x}^{(1)}+\sigma_{z}^{(1)})(\sigma_{x}^{(2)}+\sigma_{z}^{(2)})/2$ both describe dipole-dipole interactions between the spins. In the former case, the axis connecting the two spins is perpendicular to the magnetic field leading to the Zeeman term $\frac{\omega_{0}}{2} \sigma_{z}$ in \eqref{eq:Hamiltonian2}; in the latter case, the axis connecting the spins and the magnetic field are aligned under an angle of $45^{\circ}$.  
The excitation exchange interaction $\sigma_{+}^{(1)}\sigma_{-}^{(2)} + \sigma_{-}^{(1)}\sigma_{+}^{(2)}$, on the other hand, is an approximation to $\sigma_{x}^{(1)}\sigma_{x}^{(2)}$ that neglects the magnetization-changing terms $\sigma_{\pm}^{(1)}\sigma_{\pm}^{(2)}$, since $\sigma_{x}^{(1)}\sigma_{x}^{(2)} = \sigma_{+}^{(1)}\sigma_{-}^{(2)} + \sigma_{-}^{(1)}\sigma_{+}^{(2)} +  \sigma_{+}^{(1)}\sigma_{+}^{(2)} +  \sigma_{-}^{(1)}\sigma_{-}^{(2)}$.}.
The left and the central plot of Figure~\ref{fig:variationinteraction} are virtually identical. (Only the resonance widths are slightly broader in the central plot.) In particular, the symmetry-forbidden resonances remain suppressed, as $\sigma_{x}^{(1)}\sigma_{x}^{(2)}$ commutes, just like the excitation exchange interaction, with the generalized parity $\mathcal{S}_{P2}$. This is no longer the case in the right figure, where the interaction breaks the symmetry, thereby coupling all levels to each other, and triggering entanglement resonances at all driving-induced degeneracies of the uncoupled Floquet spectrum.

\paragraph*{}

To summarize, we have seen that the observed entanglement resonances are largely independent of the interaction mechanism. They can accurately be predicted by solving the \textit{single} qubit Floquet problem, Eq.~\eqref{eq:N1FloqProbl}, alone. This is by far less demanding than solving the full Floquet problem for two qubits -- a fact that becomes even more advantageous in the case of more than two qubits, that we consider in the following.

%% file: 3Qubits.tex

\section{Three Qubits and GHZ entanglement}
Based on the understanding of Floquet state entanglement of two periodically driven qubits, we extend our investigation to three qubits in the following, to see whether also in this case entanglement behaves strongly resonantly in certain parameter regions, and to pave the way for studying larger numbers of qubits.

While for two qubits, the definition of a maximally entangled state is unique, this is no longer the case for $N>2$ qubits \cite{Tichy:2011}. For three qubits, there are two inequivalent classes of maximally entangled states \cite{Dur:2000}: GHZ-entangled states, which can be transformed into
\begin{equation}\label{eq:def_GHZ}
\ket{\textrm{GHZ}} = \frac{1}{\sqrt{2}} \left( \ket{\uparrow\uparrow\uparrow} + \ket{\downarrow\downarrow\downarrow} \right)
\end{equation}
by local unitary operations, and W-entangled states, that are locally unitarily equivalent to
\begin{equation}\label{eq:def_W}
	\ket{\textrm{W}} = \frac{1}{\sqrt 3}(\ket{\downarrow\downarrow\uparrow} + \ket{\downarrow\uparrow\downarrow} + \ket{\uparrow\downarrow\downarrow}).
\end{equation}
By fixing the entanglement measure, one specifies the type of entanglement that is quantified. Here, we focus on GHZ entanglement, and use the three-tangle \cite{Coffman:2000} as entanglement measure in Eq.~\eqref{eq:def_mean_ent}, which is maximal for GHZ states and vanishes not only for separable and bi-separable states\footnote{Bi-separable states are not fully separable, but only separable with respect to a certain bipartition. An example is $\frac{1}{\sqrt 2}(\ket{\uparrow\uparrow\uparrow}+\ket{\downarrow\downarrow\uparrow})=\frac{1}{\sqrt 2}(\ket{\uparrow\uparrow}+\ket{\downarrow\downarrow})\otimes \ket{\uparrow}$.}, but also for W states.

\subsection{Identical system parameters}\label{sec:N3symmetriccase}

A natural extension of the two qubit Hamiltonian \eqref{eq:Hamiltonian2} to three qubits is
\begin{equation}\label{eq:Hamiltonian3}
H(t) = \sum_{n=1}^{3} \left( \frac{{\omega_0}}{2}  \sigma_{z}^{(n)} + f(t) \sigma_x^{(n)} \right)  + H_\textrm{qq},
\end{equation}
\begin{equation}\label{eq:HqqN3}
	H_\textrm{qq} = C \sum_{n < m}^{3}\left( \sigma_{+}^{(n)} \sigma_{-}^{(m)} + \sigma_{+}^{(n)} \sigma_{-}^{(m)}\right).
\end{equation}
Here, the interaction $H_\textrm{qq}$ couples only two qubits at a time. This is a reasonable model, since three qubit interaction is typically much weaker than pairwise interaction (e.g., in resonant excitation exchange between Rydberg atoms \cite{Vogt:2007} or chromophores \cite{Sarovar:2010}), or it is the only relevant interaction by construction (think of engineered systems, such as inductively coupled superconducting qubits \cite{Neeley:2010}). On the other hand, the assumption of all qubits to have identical coupling strength $C$ and energy splitting $\omega_0$ is somewhat artificial and often violated in reality; we discuss perturbations of this idealized situation in the next section.

For the time being, however, $H(t)$ is invariant under cyclic permutation of the qubits. We exploit this three-fold symmetry to divide the Floquet states into three decoupled symmetry classes, similar in spirit to the two qubit case, where we split off the one-dimensional antisymmetric subspace.
Again, we take a perturbative approach in the qubit-qubit interaction, and find the symmetric Floquet states of three non-interacting, driven qubits, analogously to Eq.~\eqref{eq:noninteractingFS}, to be
\begin{eqnarray}\label{eq:noninteractingFS_N3}
	\ket{\Phi_1(t)} & = &\ket{\phi_+(t)}   \ket{\phi_+(t)}   \ket{\phi_+(t)}, \\ \nonumber
	\ket{\Phi_2(t)} & = &\ket{\phi_-(t)}   \ket{\phi_-(t)}   \ket{\phi_-(t)}, \\ \nonumber
	\ket{\Phi_3(t)} & = &\frac{1}{\sqrt{3}} ( \ket{\phi_+(t)}   \ket{\phi_+(t)}   \ket{\phi_-(t)} \\ \nonumber
				& & + \ket{\phi_+(t)}   \ket{\phi_-(t)}   \ket{\phi_+(t)}\\ \nonumber
				& & + \ket{\phi_-(t)}   \ket{\phi_+(t)}   \ket{\phi_+(t)} ) , \textrm{ and} \\ \nonumber
	\ket{\Phi_4(t)} & = &\frac{1}{\sqrt{3}} ( \ket{\phi_-(t)}   \ket{\phi_-(t)}   \ket{\phi_+(t)}\\ \nonumber
				& & + \ket{\phi_-(t)}   \ket{\phi_+(t)}   \ket{\phi_-(t)}\\ \nonumber
				& & + \ket{\phi_+(t)}   \ket{\phi_-(t)}   \ket{\phi_-(t)} ).
\end{eqnarray}
Applying the local unitary transformation $U_\phi(t)$ defined in Eq.~\eqref{eq:lu}, we can read off the entanglement properties of these states:
\begin{eqnarray*}
	[U_\phi(t)]^{\otimes 3} \ket{\Phi_1(t)} = & \ket{\uparrow\uparrow\uparrow} \\
	{[}U_\phi(t) ]^{\otimes 3} \ket{\Phi_2(t)} = & \ket{\downarrow\downarrow\downarrow} \\
	{[}U_\phi(t) ]^{\otimes 3} \ket{\Phi_3(t)} = & \frac{1}{\sqrt 3}(\ket{\uparrow\uparrow\downarrow} + \ket{\uparrow\downarrow\uparrow} + \ket{\downarrow\uparrow\uparrow}) = \ket{\textrm{W}'} \\
	{[}U_\phi(t) ]^{\otimes 3} \ket{\Phi_4(t)} = & \frac{1}{\sqrt 3}(\ket{\downarrow\downarrow\uparrow} + \ket{\downarrow\uparrow\downarrow} + \ket{\uparrow\downarrow\downarrow}) = \ket{\textrm{W}}.
\end{eqnarray*}
Both $\ket{\textrm{W}}$ and $\ket{\textrm{W}'}$ are maximally W-entangled (since the latter is transformed into the former by interchanging $\ket{\uparrow}$ and $\ket{\downarrow}$ for each qubit), and altogether we have two separable and two W-entangled Floquet states in the permutation-symmetric subspace. Hence, all Floquet states have zero GHZ entanglement in the non-interacting case. However, if two of these states are tuned into resonance by a suitable driving field, a non-vanishing particle-particle interaction will induce resonant coupling, and the corresponding Floquet states will turn into superpositions of the non-interacting states \eqref{eq:noninteractingFS_N3}, what offers the possibility for GHZ entanglement.

Besides the permutation-symmetric subspace, there is the subspace associated with the permutation eigenvalue $a=e^{+2\pi i/3}$, which contains, in the non-interacting case, the Floquet states
\begin{eqnarray*}
	& {[}U_\phi(t) ]^{\otimes 3} \ket{\Phi_5(t)} = \frac{1}{\sqrt 3}(\ket{\uparrow\uparrow\downarrow} + a^* \ket{\uparrow\downarrow\uparrow} + a \ket{\downarrow\uparrow\uparrow}) & \textrm{ and} \\
 	& {[}U_\phi(t) ]^{\otimes 3} \ket{\Phi_6(t)} = \frac{1}{\sqrt 3}(\ket{\downarrow\downarrow\uparrow} + a^*\ket{\downarrow\uparrow\downarrow} +a \ket{\uparrow\downarrow\downarrow}), &
\end{eqnarray*}
and the subspace associated with the eigenvalue $a^*=e^{-2\pi i/3}$, with Floquet states obtained from those above by exchange of $a$ and $a^*$. Hence, all unperturbed Floquet states in the non-symmetric subspaces are W states. Their superposition yields at most very poor GHZ entanglement (the three-tangle is bounded by $1/9$ in these subspaces\footnote{This statement is derived as follows: Any superposition $\alpha \ket{\Phi_5(t)} + \beta \ket{\Phi_6(t)}$ is locally unitarily equivalent to $\alpha \ket{\mathrm{W}} + \beta \ket{\mathrm{W}'}$. (The corresponding transformation, after application of ${[}U_\phi(t) ]^{\otimes 3}$, reads $e^{i \pi(\sigma_z^1-\sigma_z^2)/3}$.) The three-tangle can be explicitly evaluated for this superposition, and is $\frac{16}{9}|\alpha\beta|^4$. Hence, it takes its maximum $1/9$ for $\alpha=\beta=\frac{1}{\sqrt 2}$.\label{fn:tangle}}), and therefore we focus our discussion on the symmetric subspaces in the following, which exhibits richer phenomena.

\begin{figure}[tb]
\includegraphics[width=.45\textwidth]{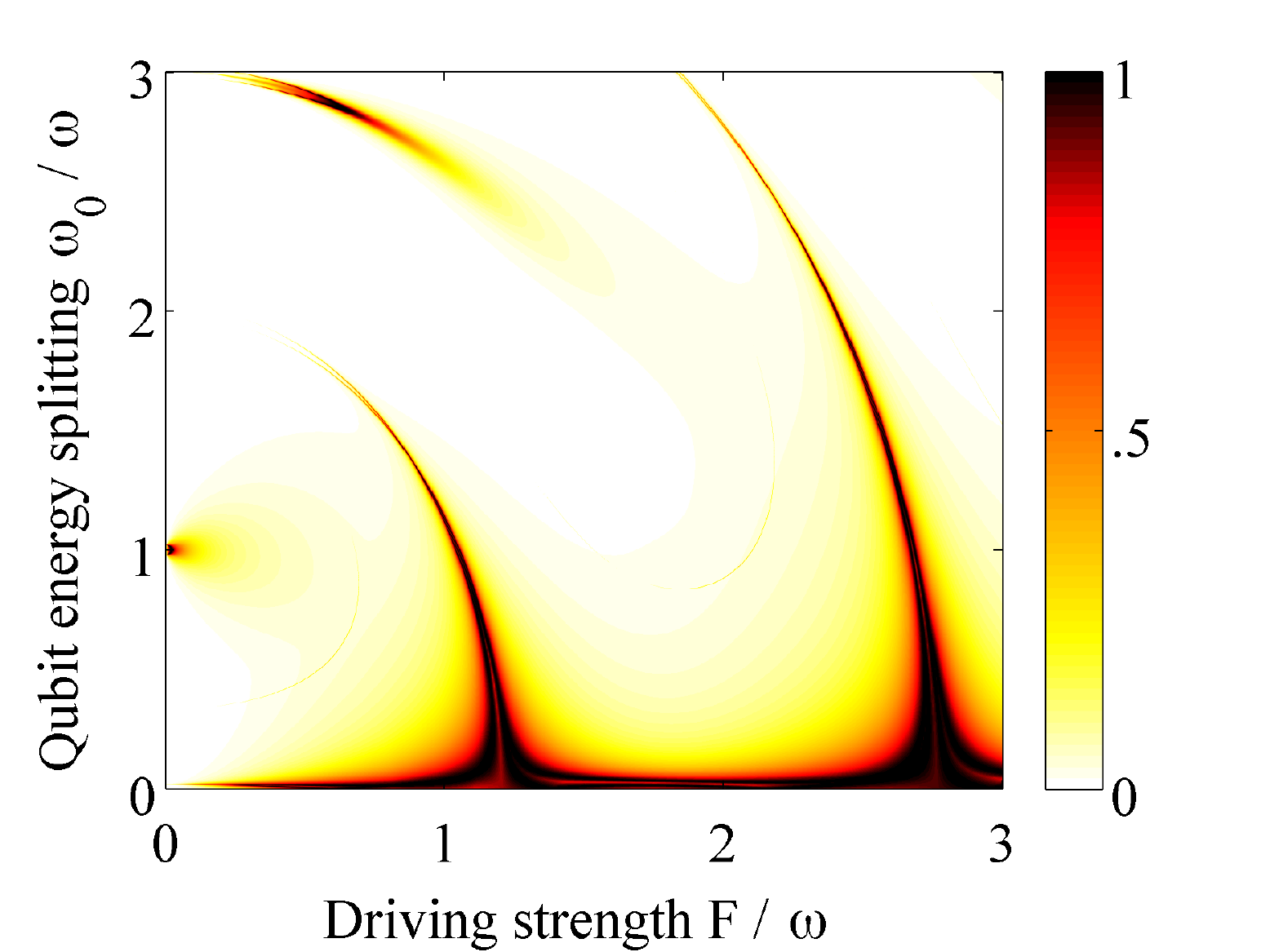}
\caption{Floquet state entanglement of three monochromatically driven qubits, as a function of driving amplitude $F$ and qubit energy splitting $\omega_{0}$, for weak interaction between the qubits ($C=0.02\,\omega$). The Floquet state with the \textit{highest} amount of GHZ entanglement is shown, in contrast to the previous figures for two qubits, in which the entanglement $\mathcal{E}_1$ of the least entangled state is of interest. Entanglement resonances appear exactly in the same parameter regions as in the two qubit case.}
\label{fig:N3_2Dplot}
\end{figure}

The results for weak interaction strength $C$ and monochromatic driving, $f(t)=F\cos(\omega t)$, are shown in Figure~\ref{fig:N3_2Dplot}. Only the permutation-symmetric Floquet state with the highest amount of entanglement is shown; i.e., at a given position in the parameter plane, we maximize over all $\mathcal{E}_{i}$.
We observe vanishing entanglement in most parts of the parameter plane, but resonant behavior in the same areas as in the two qubit case.
Accordingly, the phenomenology can be explained in the framework that was established in Sec.~\ref{sec:N2}: In the absence of interaction, none of the Floquet states exhibits any GHZ entanglement, as reasoned above. In the presence of weak interaction, the character of Floquet states is not severely altered, unless the quasi-energies of the non-interacting system are near-degenerate. For three qubits, the unperturbed quasi-energies are $\varepsilon_1=3\mu$, $\varepsilon_2=-3\mu$, $\varepsilon_3=\mu$, and $\varepsilon_4=-\mu$. Hence, whenever the degeneracy condition $\mu = n \omega/4$ of Eq.~\eqref{eq:N2degencond}, which was derived for two qubits, is fulfilled, $\varepsilon_1$ and $\varepsilon_4$ (and likewise $\varepsilon_2$ and $\varepsilon_3$) are degenerate. As in the two qubit case, the generalized parity suppresses the coupling for odd $n$; but for even $n$, the interaction matrix element $c_{14}$ is finite, and an avoided crossing opens up under finite qubit-qubit interaction. In the vicinity of the avoided crossing, the Floquet states turn into superpositions of the unperturbed states, $\alpha\ket{\Phi_1(t)} + \beta\ket{\Phi_4(t)}$, with $|\alpha|^2+|\beta|^2=1$. As one sweeps through the anti-crossing, $|\alpha|$ and $|\beta|$ continuously vary between 0 and 1, and at a certain point one always has the particular superposition
\begin{eqnarray}
	& \sqrt\frac{1}{4}\ket{\Phi_1(t)} \pm \sqrt\frac{3}{4}\ket{\Phi_4(t)} \\ \nonumber
	& = {[}U^\dagger_\phi(t) ]^{\otimes 3} \left( \sqrt\frac{1}{4}\ket{\uparrow\uparrow\uparrow} \pm \sqrt\frac{3}{4}\ket{W} \right) \\ \nonumber
	& =  {[}U^\dagger_\phi(t) ]^{\otimes 3} \left( \ket{\uparrow\uparrow\uparrow} \pm \ket{\uparrow\downarrow\downarrow} \pm \ket{\downarrow\uparrow\downarrow} \pm \ket{\downarrow\downarrow\uparrow} \right) /2.
\end{eqnarray}
This is an exact GHZ state, what can be explicitly checked by applying the Hadamard transformation $R= e^{-i \pi \sigma_y /4}$ to every qubit. Therefore, Floquet state entanglement becomes maximal at the avoided crossing of $\varepsilon_1$ and $\varepsilon_4$. Hence, the degeneracy condition $\mu = n \omega/4$, that describes the position of resonances in the two qubit case, also gives the position of GHZ entanglement resonances of three qubits.

\begin{figure}[tb]
\includegraphics[width=0.55\textwidth]{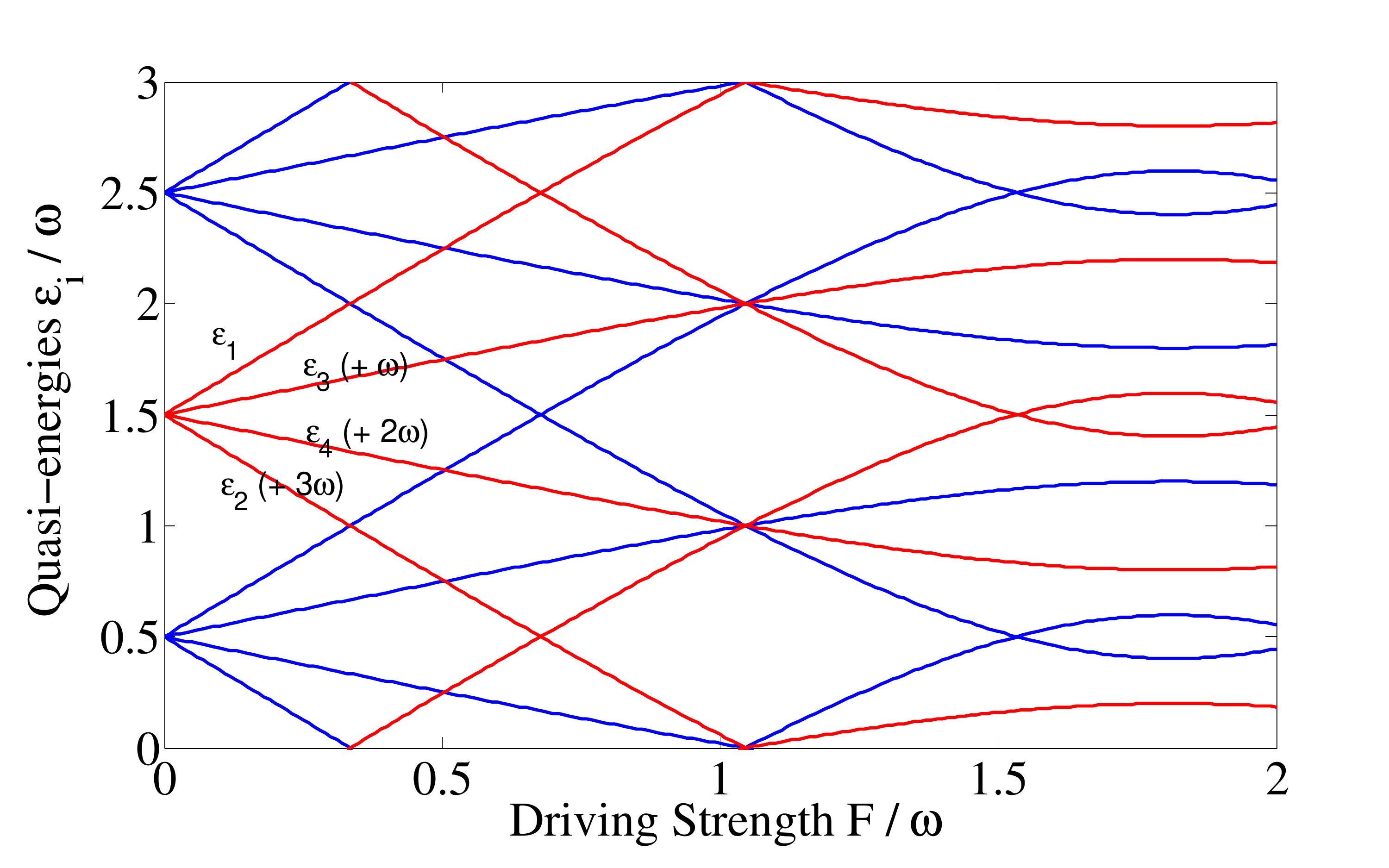}
\caption{Level structure of three non-interacting, driven qubits at $\omega_0=\omega$. For more details, see caption of the respective two qubit plot in Figure~\ref{fig:N2symmplot}.}\label{fig:N3symmplot}
\end{figure}

\begin{figure}[tb]
\includegraphics[width=0.45\textwidth]{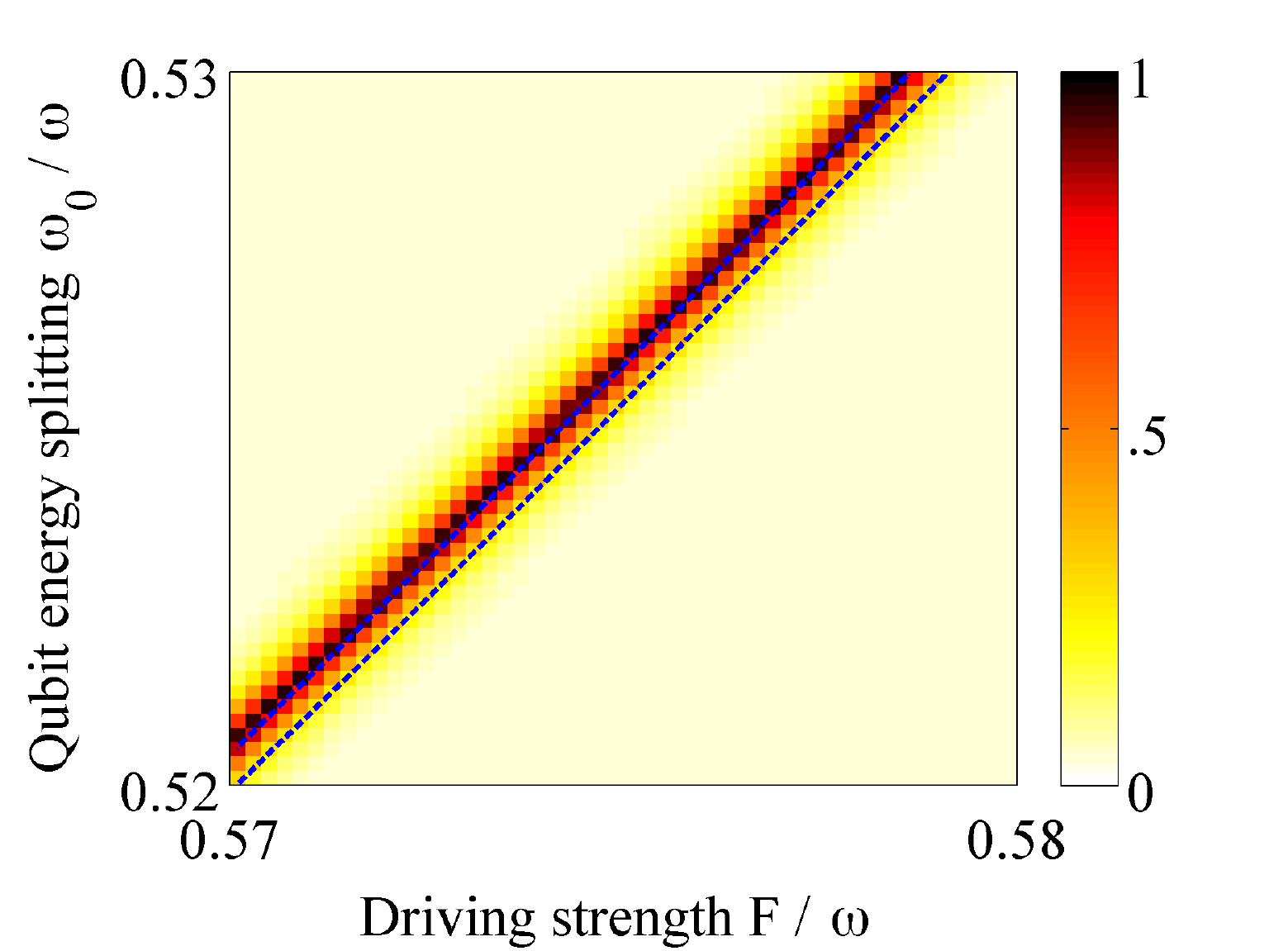}
\caption{Magnification of a small area of Figure~\ref{fig:N3_2Dplot}. A narrow entanglement resonance appears, triggered by an avoided crossing of $\varepsilon_1$ and $\varepsilon_2$. The degeneracy condition for these levels is given by Eq.~\eqref{eq:degencond_narrow}. Inside the corridor enclosed by the dashed blue lines, this condition is fulfilled for $n=2$, up to a finite tolerance $C^2$, reflecting the second order character of the coupling between the participating levels. The slight mismatch between predicted and actual position of the resonance is due to the fact that the interaction does not only open up avoided crossings, but also slightly shifts their position from the crossing of the unperturbed levels \cite{Shirley:1965}.}
\label{fig:N3_2Dzoom}
\end{figure}

As depicted in Figure~\ref{fig:N3symmplot}, there are more level crossings in the unperturbed spectrum of three qubits. The reason why they do not cause entanglement resonances in Figure~\ref{fig:N3_2Dplot} is not obvious. E.g., at the center of an avoided crossing of $\varepsilon_1$ and $\varepsilon_2$, we expect the Floquet states to turn into
\begin{equation}
	\frac{1}{\sqrt 2} \left(\ket{\Phi_1(t)} \pm \ket{\Phi_2(t)} \right)  = {[}U^\dagger_\phi(t) ]^{\otimes 3} \frac{1}{\sqrt 2} \left(\ket{\uparrow\uparrow\uparrow} \pm \ket{\downarrow\downarrow\downarrow} \right),
\end{equation}
which are maximally GHZ entangled. The degeneracy condition for $\varepsilon_1$ and $\varepsilon_2$ is
\begin{equation}\label{eq:degencond_narrow}
\mu = n \frac{\omega}{6} \qquad (n\in\mathbb{N}_0).
\end{equation}
As shown in Figure~\ref{fig:N3_2Dzoom}, entanglement resonances do indeed occur in parameter regions where condition \eqref{eq:degencond_narrow} is fulfilled for even $n$ (the odd resonances are again symmetry-suppressed). But, given the scale of Figure~\ref{fig:N3_2Dzoom}, the width of these resonance is extraordinarily small. This is why they are not visible on the scale of Figure~\ref{fig:N3_2Dplot}. The reason for this is the relevant coupling matrix element,
\begin{equation}\label{eq:CmatrixelN3}
	c_{12} = \frac{\omega}{2\pi} \int_0^{2\pi/\omega} dt \, \bra{\phi_+(t)}^{\otimes 3} H_\textrm{qq} \ket{\phi_-(t)}^{\otimes 3},
\end{equation}
which vanishes exactly for any $H_\textrm{qq}$ that mediates at most two-particle interaction, like the qubit-qubit interaction considered here, cf.~Eq.\eqref{eq:HqqN3}. (This follows from the Floquet state orthogonality $\braket{\phi_+(t)|\phi_-(t)}=0$, which implies $\bra{\phi_+(t)}^{\otimes 3} (X\otimes Y \otimes \mathbbm{1}) \ket{\phi_-(t)}^{\otimes 3}=0$ for arbitrary single particle operators $X$ and $Y$.)
Hence, $\ket{\Phi_1(t)}$ and $\ket{\Phi_2(t)}$ couple only to second order, via the intermediate states $\ket{\Phi_3(t)}$ and $\ket{\Phi_4(t)}$. Therefore, the width of their avoided crossing scales as $C^2$, which explains the sharpness of the resonance in Figure~\ref{fig:N3_2Dzoom}.

Finally, avoided crossings between $\varepsilon_3$ and $\varepsilon_4$ result in Floquet states that are locally unitarily equivalent to a superposition of $\ket{\textrm{W}}$ and $\ket{\textrm{W}'}$. As discussed for the non-symmetric permutation subspaces above (see footnote on p. \pageref{fn:tangle}), such a superposition bears only little GHZ entanglement. In addition, the transition matrix element $c_{34}$ vanishes exactly, so that the avoided crossing between these two levels opens only in second order. For both these reasons, entanglement resonances between $\varepsilon_3$ and $\varepsilon_4$ are not detected on the scale of Figure~\ref{fig:N3_2Dplot}.

\subsection{Non-identical system parameters}\label{sec:N3disordercase}

\begin{figure*}[tb]
\includegraphics[width=.45\textwidth]{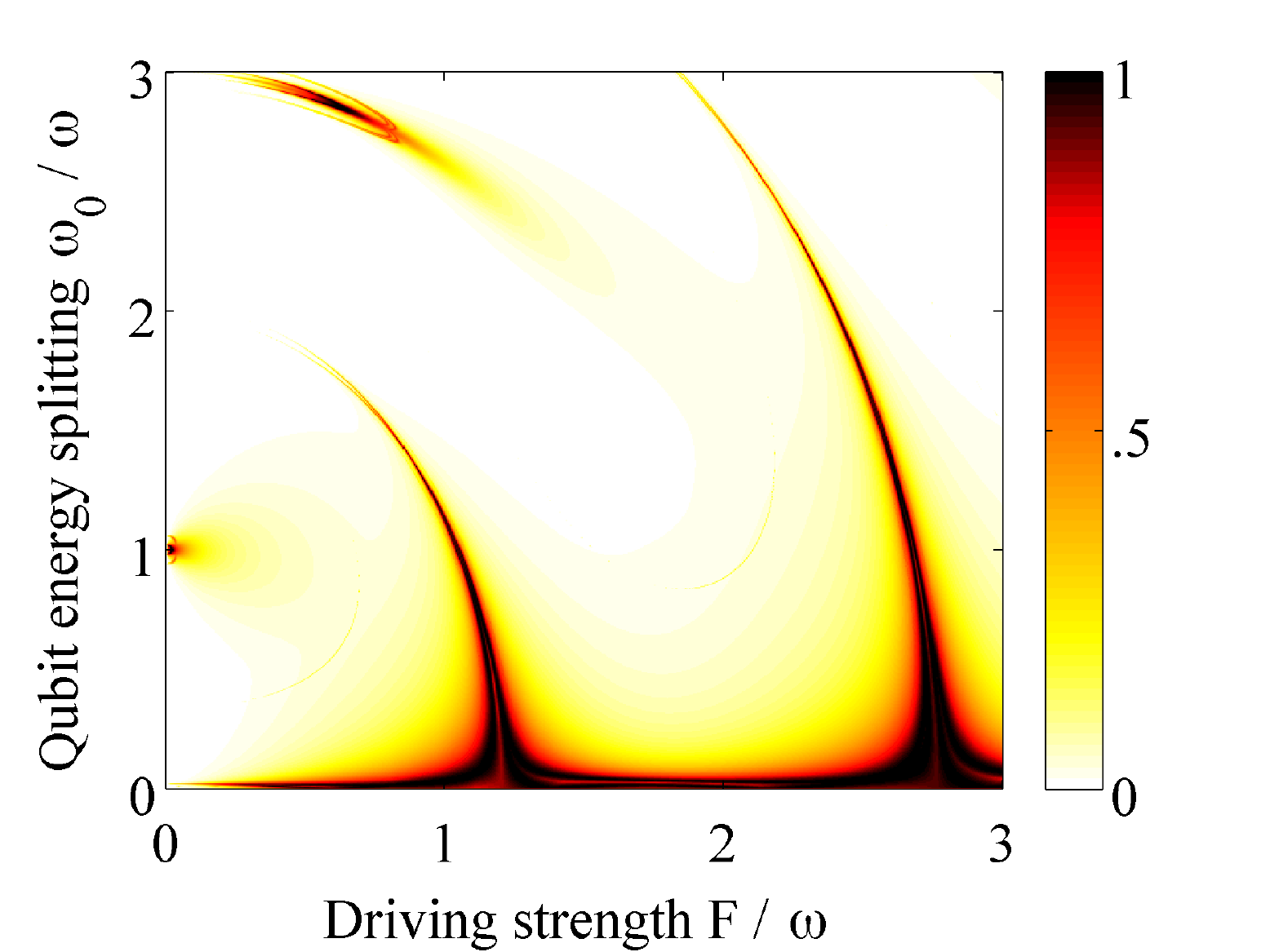}
\includegraphics[width=.45\textwidth]{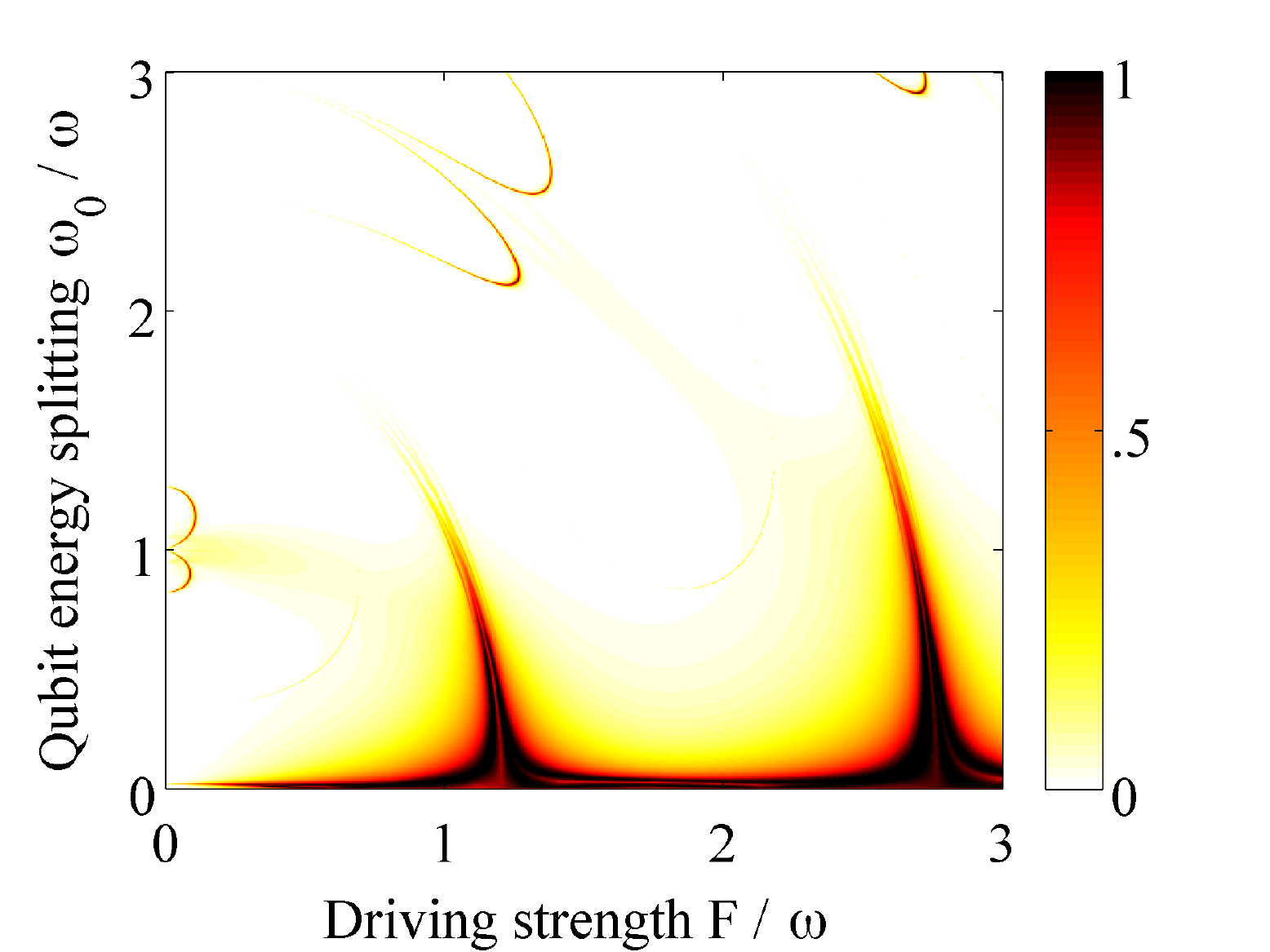}
\caption{Impact of non-identical qubit parameters on the Floquet state entanglement of three monochromatically driven, weakly coupled qubits. As in Figure~\ref{fig:N3_2Dplot}, only the Floquet state with the highest amount of GHZ entanglement is shown. In the left panel, the interaction strength $C$ between different pairs of qubits varies in a range of $\pm 10\%$. This modification does not significantly affect entanglement resonances. In the right panel, the energy splitting $\omega_{0}$ is different for each qubit, again with a spread of $\pm 10\%$. This affects the resonances in a modest, but discernible way.}
\label{fig:N3disorder}
\end{figure*}

The Hamiltonian considered so far, Eq.~\eqref{eq:Hamiltonian3}, is invariant under permutation of qubits. This assumption allowed us to reduce the number of relevant states, by focussing on the symmetric subspace. In particular, we assumed that all qubits have equal energy splitting $\omega_0$, and that the strength $C$ of the qubit-qubit interaction is identical for all pairs of qubits. In certain cases (e.g., in engineered quantum systems), this can be a valid assumption; but in most naturally occurring situations, one often rather has slightly different parameters for each qubit.

Therefore, we study a slight deviation from identical system parameters in Figure~\ref{fig:N3disorder}. The left panel shows the influence of individual coupling strengths $C$ for each pair of qubits. Each term in the qubit-qubit interaction is of different strength now,
\begin{equation*}
	H_\textrm{qq} = C \sum_{n < m}^{3} \alpha_{nm} \left( \sigma_{+}^{(n)} \sigma_{-}^{(m)} + \sigma_{+}^{(n)} \sigma_{-}^{(m)}\right),
\end{equation*}
reflected by the weighting factors $\alpha_{nm}$, which are chosen as $\{ \alpha_{12}=0.9, \alpha_{23}=1.0, \alpha_{31}=1.1\}$. The impact of such modification is not apparent, and the entanglement resonances are practically not affected (compare Figure~\ref{fig:N3_2Dplot}). This is consistent with our discussion in Sec.~\ref{sec:interactionvariation}, where we reasoned why a change of the interaction term only marginally influences the resonances. The same argument holds here: Since the role of $H_\textrm{qq}$ is only to open the avoided crossings, a variation of its strength affects only the width of resonances, not their shape.

The situation is different in right panel of Figure~\ref{fig:N3disorder}: Here, the qubit energy splittings $\omega_0$ are individualized by putting
\begin{equation}\label{eq:Hamiltonian3dis}
	H(t) = \sum_{n=1}^{3} \left( \beta_{n} \frac{{\omega_0}}{2}  \sigma_{z}^{(n)} + f(t) \sigma_x^{(n)} \right)  + H_\textrm{qq},
\end{equation}
with weighting factors $\{ \beta_1=0.9, \beta_2 =1.0, \beta_3 =1.1\}$. Since this modification influences the single qubit quasi-energy $\mu$, the position of level crossings in the unperturbed spectrum is affected, and, accordingly, the shape of the resonances may change. Yet, the two dominant resonances Figure~\ref{fig:N3disorder} retain their overall shape and width, and only the smaller resonances split into ``twin arches''.

In conclusion, we find that a modest deviation from the assumption of identical system parameters for each qubit does not significantly alter the phenomenology of entanglement resonances.

%% file: NQubits.tex

\section{$N>3$ qubits and general discussion of entanglement resonances}\label{sec:Uniqueness}

We have seen that the phenomenon of entanglement resonance is neither restricted to two qubits, nor to a particular driving profile $f(t)$ or qubit-qubit interaction $H_\textrm{qq}$. In the following, we extend our analysis to the general case of $N>3$ weakly interacting, driven qubits.

\subsection{Identical system parameters}

In the symmetric case of identical qubit parameters, the $N$-qubit Hamiltonian we want to study reads
\begin{equation}\label{eq:HamiltonianN}
H(t) = \sum_{n=1}^{N} \left( \frac{{\omega_0}}{2}  \sigma_{z}^{(n)} + f(t) \sigma_x^{(n)} \right)  + H_\textrm{qq}.
\end{equation}
As discussed for $N=3$ qubits in Sec.~\ref{sec:N3symmetriccase}, $H_\textrm{qq}$ is typically restricted to two-body interactions. As an example, we study again the excitation exchange interaction
\begin{equation}\label{eq:HqqNN}
H_\textrm{qq} = C \sum_{n < m}^{N}\left( \sigma_{+}^{(n)} \sigma_{-}^{(m)} + \sigma_{+}^{(n)} \sigma_{-}^{(m)}\right),
\end{equation}
and introduce the collective spin operator
\begin{equation}\label{eq:collectivespin}
	\vec{J} = \frac{1}{2} \sum_{n=1}^{N} \vec{\sigma}^{(n)}  \qquad (\vec{\sigma} ^{(n)} =\{\sigma ^{(n)}_x,\sigma ^{(n)}_y,\sigma ^{(n)}_z\}).
\end{equation}
With $J_\pm=(J_x \pm i J_y)/2$, the Hamiltonian rewrites
\begin{equation}
H(t) = \omega_0  J_z + 2 f(t) J_x  + \frac{C}{2} (J_{+}J_{-}+J_{-}J_{+}-N).
\end{equation}
Since $H(t)$ commutes with $\vec{J}^2$, the dynamics conserves the total spin, with values $N/2, N/2-1, \dots, 0,$ for even $N$, or $N/2, N/2-1, \dots, 1/2,$ for odd $N$. The subspace with maximal total spin has dimension $N+1$, and contains all states that are symmetric under cyclic permutation of the qubits. As in the previous chapters for two and three qubits, we focus on this symmetric subspace in our discussion of Floquet state entanglement. This is not a severe limitation, since it is 
reasonable to assume that an interacting system of identical qubits is initially in a symmetric state, e.g., the de-excited state $\ket{\downarrow}^{\otimes N}$, and can hence only explore the symmetric subspace under the dynamics induced by Hamiltonian~\eqref{eq:collectivespin}.

Introducing the Dicke states \cite{Dicke:1954,Gross:1982}
\begin{eqnarray}\label{eq:Dickestate}
	\ket{N,m}\equiv \left[ \binom N m \right]^{-1/2} J_{+}^{m} \ket{\downarrow}^{\otimes N}, \qquad m\in\{0,\dots,N\},
\end{eqnarray}
as a basis of the symmetric subspace ($m$ corresponds to the number of ``excitations'', i.e., to the number of qubits in the spin-up state), the symmetric Floquet states of the non-interacting system, $C=0$, read
\begin{equation}\label{eq:noninteractingFS_NN}
	\ket{\Phi_i(t)} = [U^\dagger_\phi(t)]^{\otimes 3} \ket{N,i}.
\end{equation}
$U_\phi(t)$ is defined via the single qubit Floquet states, see Eq.~\eqref{eq:lu}. Consequently, the entanglement properties of $\ket{\Phi_i(t)}$ are equivalent to those of $\ket{N,i}$. The latter are studied in detail in \cite{Bastin:2009,Hayashi:2008,Markham:2011}. One finds that $\ket{N,0}=\ket{\downarrow}^{\otimes N}$ and $\ket{N,N}=\ket{\uparrow}^{\otimes N}$ are the only separable states, and that the remaining Dicke states belong to distinct SLOCC classes (at least for $i\le N/2$ or $i\le(N+1)/2$, respectively, since $\ket{N,i}$ and $\ket{N,N-i}$ are connected by the local operation that exchanges $\uparrow$ and $\downarrow$ labels, and therefore have equivalent entanglement properties). All these distinct classes have W character \cite{Bastin:2009}.

In the presence of a weak qubit-qubit interaction, the Floquet states $\ket{\Phi_i(t)}$ are not altered significantly, as long the quasi-energies $\varepsilon_i$ of the non-interacting case are far from degeneracy. On the other hand, when two levels $\varepsilon_i$ and $\varepsilon_j$ cross under variation of some parameter (e.g., the driving amplitude $F$ or the driving frequency $\omega$, which are usually easily tunable), the interaction can lift this degeneracy and induce an avoided crossing, under conditions discussed below. By sweeping the parameters through such an avoided crossing, any superposition $\alpha \ket{\Phi_i(t)} + \beta \ket{\Phi_j(t)}$ will become a Floquet state of $H(t)$ at some point. This way, Floquet states can be ``designed'' to be locally unitarily equivalent to any superposition of two Dicke states $\ket{N,i}$ and $\ket{N,j}$, simply by tuning the respective quasi-energies $\varepsilon_i$ and $\varepsilon_j$ into resonance. This opens up the possibility of achieving Floquet states entanglement of many more classes, compared to the W classes the bare Dicke state belong to. E.g., the balanced superposition of $\ket{N,0}$ and $\ket{N,N}$ results in the $N$-qubit GHZ state. For four qubits, $\sqrt\frac{1}{3}\ket{4,0}+ \sqrt\frac{2}{3}\ket{4,3}$ bears ``T-entanglement'', which is SLOCC inequivalent to ``single-excitation W entanglement'' (found in $\ket{4,1}$ and $\ket{4,3}$), ``two-excitation W'' (found in $\ket{4,2}$), and GHZ entanglement \cite{Markham:2011}.

In order to design Floquet states in this manner, one needs to determine the driving parameters that tune $\varepsilon_i$ and $\varepsilon_j$ into resonance. Since the non-interacting quasi-energies are $\varepsilon_i=[(N-2i)\mu]\mod \omega$, with the single qubit quasi-energy $\mu$ defined in Eq.~\eqref{eq:N1FloqProbl}, the resonance condition reads
\begin{equation}\label{eq:NNdegencond}
	\mu = n \frac{\omega}{2(i-j)}, \qquad n\in \mathbb{N}_0.
\end{equation}
In order to find the right parameters for the desired level crossing, if therefore suffices to solve the Floquet problem of the \textit{single} qubit Hamiltonian $h(t)$, and to identify parameters which fulfill the resonance condition. This is by far simpler than solving the Floquet problem of the full, interacting $N$-qubit Hamiltonian $H(t)$.

It remains to determine how strongly the non-interacting Floquet levels are coupled by the qubit-qubit interaction: If two levels are not coupled at all, no avoided crossing occurs between them, and no superposition of Dicke states can be created in the fashion sketched above. Furthermore, if levels couple very weakly, their avoided crossing is small, and the driving parameters have to be tuned very precisely to establish the desired superposition, a task that becomes experimentally unfeasible below a certain threshold. Analogously to the case of two and three qubits in Eqs.~\eqref{eq:Cmatrixel} and \eqref{eq:CmatrixelN3}, the coupling strength is determined in first order by the interaction operator Floquet matrix element
\begin{equation}\label{eq:CmatrixelNN}
	c_{ij} = \frac{\omega}{2\pi} \int_0^{2\pi/\omega} dt \, \bra{\Phi_i(t)} H_\textrm{qq} \ket{\Phi_j(t)}.
\end{equation}
If $H_\textrm{qq}$ mediates only two-qubit interactions, $c_{ij}$ vanishes for $|i-j|>2$. Therefore, $\ket{\Phi_i(t)}$ and $\ket{\Phi_{i+2m}(t)}$ are only coupled at $m$-th order, and the width of their avoided crossing scales like $C^m$. Hence, the ability to superimpose these states by tuning into the avoided crossing rapidly decreases with increasing $m$. The creation of GHZ entanglement by superimposing $\ket{\Phi_0(t)}$ and $\ket{\Phi_N(t)}$ is particularly affected by this issue, since there $m=(N+1)/2$. For three qubits, this leads to the second order character of the entanglement resonance shown in Figure~\ref{fig:N3_2Dzoom}.

\subsection{Non-identical system parameters}

The analysis in terms of the collective spin operator $\vec{J}$ and the symmetric Dicke states $\ket{N,i}$ above was possible because all $N$ qubits were assumed to be identical in the Hamiltonian \eqref{eq:HamiltonianN}. Now, we drop this assumption and introduce individual parameters for each qubit:
\begin{equation}\label{eq:HamiltonianNdisorder}
H(t) = \sum_{n=1}^{N} \left( \frac{\omega_0^{(n)}}{2}  \sigma_{z}^{(n)} + f^{(n)}(t) \sigma_x^{(n)} \right)  + H_\textrm{qq},
\end{equation}
\begin{equation}
H_\textrm{qq} = \sum_{n < m}^{N} C^{(n,m)}\left( \sigma_{+}^{(n)} \sigma_{-}^{(m)} + \sigma_{+}^{(n)} \sigma_{-}^{(m)}\right)
\end{equation}
For three qubits, we have phenomenologically studied the influence of such modifications on Floquet state entanglement in Sec.~\ref{sec:N3disordercase}.


In general, individual parameters for each qubit break the permutation invariance of $H(t)$. Generically, this leads to completely different entanglement properties of Floquet states than those discussed above for the symmetric case.
This is due to the single qubit problem, Eq.~\eqref{eq:N1FloqProbl}, being different for each qubit now, and leading to individual single qubit Floquet states $\ket{\phi^{(n)}_\pm(t)}$ and quasi-energies $\mu^{(n)}$. Then, Floquet states of the non-interacting system are product states
\begin{equation}\label{eq:noninteractingFS_NNdisorder}
 \ket{\Phi_{\vec{s}}(t)} = \bigotimes_{n=1}^{N} \ket{\phi^{(n)}_{s_n}(t)},
\end{equation}
labeled by a string $\vec{s}$ of $N$ plus or minus signs, and have quasi-energies
\begin{equation}
 \varepsilon_{\vec{s}} = \sum_{n=1}^N s_n \mu^{(n)}.
\end{equation}
Let us first explain how the case of identical parameters fits into this picture: There, due to $\mu^{(n)}\equiv\mu$, quasi-energies $\varepsilon_{\vec{s}}$ and $\varepsilon_{\vec{p}}$ are degenerate, if $\vec{s}$ and $\vec{p}$ contain an equal number of plus signs. The product states \eqref{eq:noninteractingFS_NNdisorder} are not a clever choice of basis in these degenerate subspaces: Since we later want to include a permutation-invariant perturbation $H_\textrm{qq}$, the appropriate basis is rather the one that diagonalizes the permutation operator. This way, focussing on one particular permutation class (e.g., the symmetric one), all ambiguities are eliminated: We have quasi-energies $\varepsilon_{\vec{s}}=(N-i)\mu$ (where $i$ is the number of plus signs in $\vec{s}$), and the symmetrized Floquet states of Eq.~\eqref{eq:noninteractingFS_NN}.

Such systematic degeneracies are absent in the case of non-identical parameters. Generically, the only basis of the unperturbed system are the product states $\ket{\Phi_{\vec{s}}(t)}$ of Eq.~\eqref{eq:noninteractingFS_NNdisorder}. Restricting the qubit-qubit coupling $H_\textrm{qq}$ to two-qubit interaction again, the product states $\ket{\Phi_{\vec{s}}(t)}$ and $\ket{\Phi_{\vec{p}}(t)}$ are coupled by $H_\textrm{qq}$ (in first order) only if $\vec{s}$ and $\vec{p}$ differ in at most two entries. Then, in the vicinity of the avoided crossing of $\varepsilon_{\vec{s}}$ and $\varepsilon_{\vec{p}}$, the two corresponding Floquet states turn into the superposition of $\ket{\Phi_{\vec{s}}(t)}$ and $\ket{\Phi_{\vec{p}}(t)}$. Since all $N-2$ subsystems corresponding to identical entries of $\vec{s}$ and $\vec{p}$ can be factored out in this superposition, only bipartite entanglement between the remaining two qubits is generated at this entanglement resonance.

The limitation to bipartite entanglement for the first order entanglement resonances can only be overcome if more than two levels meet in an avoided crossing. Apparently, this is the case in Figure~\ref{fig:N3disorder}, where plenty of GHZ entanglement is present. This can be ascribed to the fact that only a slight deviation from identical parameters was studied there, which retain some systematic degeneracies of the symmetric case, in which multipartite entanglement is the norm rather than an exception.

%% file: Conclusion.tex

\section{Summary and conclusion}

In this paper, we considered weakly interacting qubits that are coherently driven by an external, time-periodic field. This situation is realized in many experiments with quantum-mechanical two-level systems, such as trapped ions \cite{Kim:2010}, superconducting qubits \cite{McDermott:2005}, Nitrogen-vacancy centers in diamond \cite{Neumann:2010}, or cold Rydberg atoms \cite{Gallagher:2008mz}. By employing the Floquet picture, we are able to analyze parameter ranges beyond the rotating wave approximation (RWA), i.e., strong and off-resonant driving. The main goal was to study the entanglement of the Floquet states of such systems.

For two qubits, we first analyzed the case of identical parameters for both qubits, and found that two of the three Floquet states of the symmetric subspace are only entangled in certain regions of the parameter space. The occurrence of these entanglement resonances was studied in more detail in the following: An explanation was given for the emergence of the resonances, based on a perturbative treatment of the qubit-qubit interaction, and the avoided crossings induced by this interaction. This led to the necessary, but not sufficient condition \eqref{eq:N2degencond} for entanglement resonance that nicely describes the shape of resonances in parameter space.
Based on this understanding, we studied different time-dependencies of the driving field (monochromatic, bi-chromatic, saw-tooth), different interaction mechanisms, and the influence of individual parameters for both qubits, and found that all these variations can be well accounted-for by our theoretical framework.

For three qubits, we focussed on the GHZ entanglement class and again found entanglement resonances in parameter space. An explanation of this observation was given along the same lines as in the two qubit case. Finally, we generalized our reasoning to $N>3$ qubits, for both identical and non-identical system parameters.

Altogether, the main results of this work are
\begin{enumerate}
\item the finding that maximally entangled Floquet states exist at avoided crossings in the Floquet spectrum of weakly interacting, periodically driven quantum systems. While a similar phenomenon has been reported in weakly interacting systems under \textit{static} forcing \cite{Bruss:2005,Karthik:2007}, the advantage of the here proposed control scheme is that alternating control fields often provide the simplest way to control a quantum system (think of, e.g., trapped ions \cite{Leibfried:2003}, cold atoms \cite{Zenesini:2009}, or color centers in diamonds \cite{Wrachtrup:2006}). In addition, control via time-periodic fields is more versatile, since the quantum system can be addressed through a multitude of side-bands, and it can lead to increased coherence times \cite{Timoney:2011}, suggesting that the entanglement of Floquet states might be more robust against decoherence than the entanglement of eigenstates of an undriven system. Verification of this conjecture is a promising perspective for future studies.
\item a simple prediction scheme for the positions of the avoided crossings (and therefore of entanglement resonances) in parameter space, based on the degeneracy condition \eqref{eq:N2degencond}. Since this condition involves only the quasi-energy $\mu$ of a \textit{single} driven qubit, it does not increase in complexity as the number of qubits increases, and can even be evaluated analytically for certain driving profiles.

Notably, the prediction scheme is not limited to the analysis of Floquet state entanglement, since entanglement is not the only property that behaves critically in the vicinity of avoided level crossings: In fact, as discussed in Sec.~\ref{sec:N2}, the qubit-qubit interaction (when regarded as a perturbation to the non-interacting, periodically driven system) can change the character of Floquet states significantly only if two quasi-energies are in resonance, i.e., close to degeneracy. Hence, by tuning the driving parameters in or out of such a resonance, one effectively enhances or suppresses the interaction in a controlled way. Identifying the resonant driving parameters is therefore essential for the analysis of \textit{any} interaction-related property of a periodically driven quantum system. Moreover, by bringing more than two levels into resonance, one can dedicatedly study the effect of three-body, four-body, etc., interaction. Therefore, the scheme presented here does not only explain the behavior of Floquet state entanglement, but more generally provides a recipe for controlling many-body interactions by periodic driving fields, e.g., in nuclear spin systems \cite{Kropf:2011} or Rydberg gases \cite{Gurian:2012}.
\end{enumerate}

\ack
S.S. acknowledges financial support by the German National Academic Foundation. A.B. acknowledges partial support through COST action MP1006. 

%% file: Appendix.tex

\section{Rule of thumb for the number of Fourier components}\label{sec:AppendixNoFouriercomp}

In this Appendix, we derive a rule of thumb for the number $M$ of Fourier components that have to be taken into account in order to appropriately describe the Floquet states of a single, monochromatically driven qubit.

In the language of Sec.~\ref{sec:Floquet_details}, we seek $M\in\mathbb{N}$ such that
\begin{equation}\label{eq:ApxM}
	 \forall k \in \mathbb{Z} \textrm{ with } |k|> M: \qquad  || \ket{\tilde \phi_i(k)} || \ll 1
\end{equation}
Here, $\ket{\tilde \phi_i(k)}$ is the $k$-th Fourier component of the Floquet states $\ket{\phi_i(t)}$ of
\begin{equation}
	h(t) = \frac{\omega_0}{2}\sigma_z + F\cos(\omega t) \sigma_x.
\end{equation}
The respective Floquet Hamiltonian $\mathbf{h}_F$ is
\begin{eqnarray}\label{eq:HF_Appendix}
	\mathbf{h}_F=\frac{1}{2}\left( \begin{matrix}
	  \ddots & & \vdots & & \\
	   & \omega_0 \sigma_z + 2 \omega & F \sigma_x & 0 &  \\
           \hdots& F \sigma_x & \omega_0 \sigma_z  & F \sigma_x &  \hdots \\
	  & 0 & F \sigma_x & \omega_0 \sigma_z - 2 \omega  &   \\
	  & & \vdots & & \ddots \\
	\end{matrix}\right)
\end{eqnarray}
in Fourier representation. Due of the generalized parity symmetry $\mathcal{S}_P$, cf. Eq.~\eqref{eq:SP}, $\mathbf{h}_F$ consists of two uncoupled blocks $\mathbf{h}^{(\pm)}_F$, according to the two parity classes \footnote{
In the notation of Sec.~\ref{sec:Floquet_details}, the block $\mathbf{h}^{(\pm)}_F$ of positive (negative) parity comprises the basis states $\ket{\uparrow}\otimes\ket{k}$ and $\ket{\downarrow}\otimes\ket{k+1}$ with even (odd) $k$.}:

\begin{eqnarray}
	\mathbf{h}^{(\pm)}_F=\frac{1}{2}\left( \begin{matrix}
	  \ddots & & \vdots & & \\
	   & \mp\omega_0  + 2 \omega  & F & 0 &  \\
           \hdots& F & \pm \omega_0  & F  &  \hdots \\
	  & 0 & F & \mp \omega_0 - 2 \omega  &   \\
	  & & \vdots & & \ddots \\
	\end{matrix}\right)
\end{eqnarray}
For $\omega_0=0$, the eigenvalues of this matrix are $j\omega$ (with $j\in\mathbb{Z}$). The $k$-th frequency component of the corresponding eigenstates $\ket{\ket{\phi_j}}$ is given by $\ket{\tilde\phi_j(k)} = J_{|j-k|}(F/\omega)$ \cite{Hartmann:2004}, with $J_m$ denoting the $m$-th order Bessel function of the first kind. Hence, $\ket{\ket{\phi_j}}$ is centered around the $j$-th frequency component, and the different eigenstates are frequency shifted versions of each other. It therefore suffices to study the number of frequency components $M$ of $\ket{\ket{\phi_0}}$: From numerics, we find that $|J_{|k|}(F/\omega)|<.05$ holds for $k \gtrapprox 2 F/\omega$, and therefore we have $M=2F/\omega$.

\begin{figure}[tb]
\includegraphics[width=0.45\textwidth]{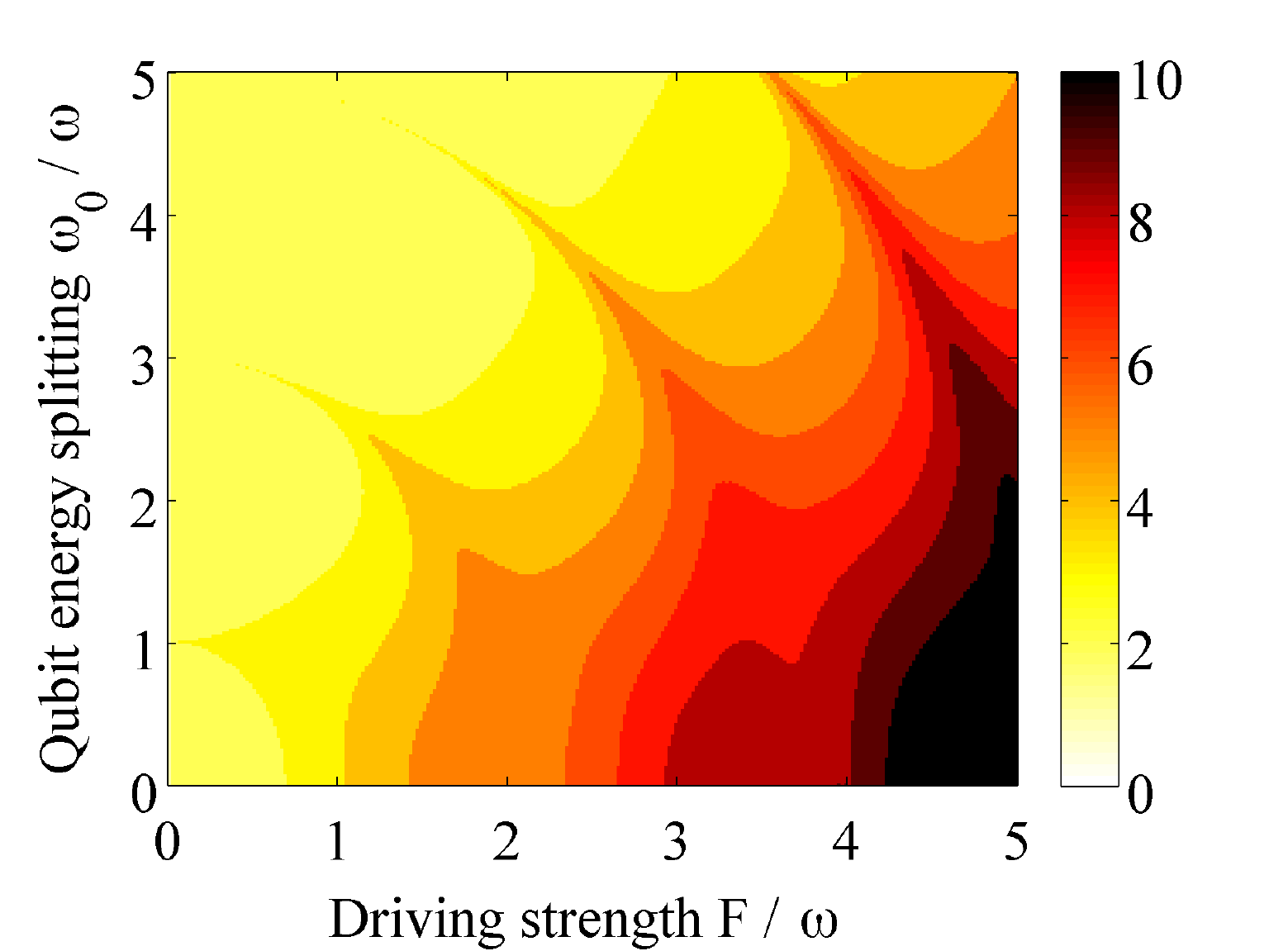}
\caption{Number of frequency components $M$ of the Floquet states of a monochromatically driven qubit, as defined in Eq.~\eqref{eq:ApxM}. The value of $M$ is indicated by a color code, and plotted as a function of driving strength $F$ and qubit level splitting $\omega_0$. At $\omega_0=0$, one has $M\approx2F/\omega$, as discussed in the text. For $\omega_0\ne0$, $M$ is always smaller than this estimate.}\label{fig:ApxM}
\end{figure}
Figure~\ref{fig:ApxM} depicts the behavior of $M$ for $\omega_0\ne0$: At any driving strength $F$, the value of $M$ for $\omega_0\ne0$ is always smaller than for $\omega_0=0$. Therefore, the expression $M=2F/\omega$, which is strictly valid for $\omega_0=0$, \textit{always} defines an upper bound of $M$, and can hence be used as a general rule of thumb.

\section{Quasi-energies for saw-tooth and $\delta$-kicked driving}\label{sec:AppendixSawtooth}

In this Appendix, we derive the exact quasi-energies of a single qubit, driven with a saw-tooth profile, or by periodic $\delta$-kicks. To the best of our knowledge, both expressions have not been derived in the literature so far.

In Sec.~\ref{ssec:N2bichromsawtooth}, we consider the saw-tooth driving profile $f(t)=F \cdot \left[\frac{t}{T} \textrm{ mod } 1 - \frac{1}{2}\right]$. It describes an external field that is periodically ramped up from $-F/2$ to $F/2$, with period $T=2\pi/\omega$. Let us consider, for the moment, an additional constant ``offset'' $F/2$, such that $f(t)=F \cdot \left[\frac{t}{T} \textrm{ mod } 1\right]$. (I.e., the driving field is ramped up from $0$ to $F$ now.) Within the time interval $[0,T)$, the evolution is thus governed by the Hamiltonian
\begin{equation}
	h(t) = \frac{\omega_0}{2} \sigma_z + \frac{Ft}{T} \sigma_x.
\end{equation}
Comparing this to the Hamiltonian
\begin{equation}
	h_\textrm{LZ}(t) = \frac{\alpha t}{2} \sigma_z + V \sigma_x
\end{equation}
of a Landau-Zener scenario \cite{Akulin:2006}, we see that $h(t)$ and $h_\textrm{LZ}(t)$ are basically identical; only the roles of $\sigma_z$ and $\sigma_x$ are interchanged. Indeed, by identifying  $V\equiv\omega/2$ and $\alpha \equiv-2F/T$, and defining the Hadamard transformation $R= e^{-i \pi \sigma_y /4}$, we have
 \begin{equation}
	h(t) = R \, h_\textrm{LZ}(t) \, R^{-1}.
\end{equation}
Since $R$ is static and unitary, quasi-energies are invariant under this transformation. To derive the quasi-energies of $h_\textrm{LZ}(t)$, we make use of the fact that its time evolution operator $U(0,t)$ is explicitly known \cite{Akulin:2006}:
\begin{eqnarray}\label{eq:ApxUt}
	& U(0,t) &=\left( \begin{matrix}
	  a(t) & -b(t) \\
	  b(t)^* & a(t)^* \end{matrix} \right) \\
\nonumber	 & a(t) &= \, _1 F_1\left(\frac{i V^2}{2 \alpha},\frac{1}{2}, \frac{-i\alpha t^2}{2} \right) e^{-i\alpha t^2 /4}\\
\nonumber	 & b(t) &= \, _1 F_1\left(\frac{1}{2} + \frac{i V^2}{2 \alpha},\frac{3}{2}, \frac{-i\alpha t^2}{2} \right) i V t \, e^{-i\alpha t^2 /4},
\end{eqnarray}
with $_{1}F_{1}$ denoting Kummer's function \cite{Abramowitz:1964}. The eigenvalues of this operator are $e^{\pm i \textrm{arccos} (\Re [a(t)])}$, $\Re$ denoting the real part; hence, its eigenphases are $\pm \textrm{arccos} (\Re [a(t)])$. The quasi-energies $\mu_\pm$ of $h_\textrm{LZ}(t)$ -- and therefore also of $h(t)$ -- are the eigenphases of $U(0,T)$, divided by $T$ \cite{Haake:1991}. Inserting the definition of $a(t)$ and the relations $V\equiv\omega/2$ and $\alpha\equiv -2F/T$, we finally have
 \begin{equation}
	\mu_\pm = \pm \frac{1}{T} \textrm{arccos} \left(\Re [\, _1 F_1\left(\frac{- i \omega_0^2 T}{16 F},\frac{1}{2}, -i F T \right) e^{i F T /2}] \right).
\end{equation}

To derive the quasi-energies in the absence of the additional offset introduced above, i.e., for $f(t)=F \cdot \left[\frac{t}{T} \textrm{ mod } 1 - \frac{1}{2}\right]$, we simply shift the time interval $[0,T)$ by half a period, since the time evolution generated by the Landau-Zener Hamiltonian $h_\textrm{LZ}(t)$ during the time interval $[-T/2,T/2)$ precisely described the dynamics of saw-tooth driving without the offset. The time evolution operator now reads
\begin{eqnarray}
\nonumber	U(-T/2,T/2) & = & U^\dagger(0,-T/2) \, U(0,T/2) \\
	& \overset{\eqref{eq:ApxUt}}{=} & \left( \begin{matrix}
	   |a|^2 - |b|^2 & 2 a^* b \\
	  -2 a b^* & |a|^2 - |b|^2 \end{matrix} \right),
\end{eqnarray}
where we used the short-hand notation $a$ and $b$ instead of $a(T/2)$ and $b(T/2)$. From the eigenphases $\pm \textrm{arccos}(|a|^2 - |b|^2)$ of this operator, we find, together with the normalization condition $|a|^2 + |b|^2=1$:
\begin{eqnarray}
			\mu_\pm &=& \pm \frac{1}{T} \textrm{arccos} \left(2 |a|^2 -1 \right) \\
\nonumber 		& = & \pm \frac{1}{T} \arccos\left[ 2 \left|\, _1 F_1\left(\frac{-i \omega_0^2 T}{16 F},\frac{1}{2}, -\frac{i F T}{4} \right) \right|^2 - 1 \right].
\end{eqnarray}

Likewise, the quasi-energies of a periodically $\delta$-kicked qubit can be obtained analytically. The driving profile is $f(t)=FT \cdot \delta(t\mod T)$, and the time evolution operator reads
\begin{eqnarray}
	& U(0,t) &= e^{-i FT \sigma_x} e^{-i \frac{\omega_0}{2} t \sigma_z} = \left( \begin{matrix}
	  a(t) & -b(t) \\
	  b(t)^* & a(t)^* \end{matrix} \right) \\
\nonumber	 & a(t) &= e^{-i \frac{\omega_0}{2} t } \cos(FT)\\
\nonumber	 & b(t) &= i\, e^{i \frac{\omega_0}{2} t } \sin(FT),
\end{eqnarray}
leading to quasi-energies
\begin{eqnarray}\label{eq:KickedExplicitQE}
\nonumber	\mu_\pm & =& \pm \frac{1}{T} \textrm{arccos} (\Re [a(T)]) \\
				&=& \pm \frac{1}{T} \arccos\left[ \cos\left(\frac{\omega_0 T}{2}\right) \cdot \cos (FT) \right].
\end{eqnarray}